\documentclass{jimis-en}
\pdfoutput=1 

\usepackage{array}
\usepackage{pgfplots}
\usepackage{graphicx}

\title{A general graph-based framework for top-N recommendation using content, temporal and trust information}
\author[*1,2,3]{Armel Jacques NZEKON NZEKO'O}
\author[1,2]{Maurice TCHUENTE}
\author[3]{Matthieu LATAPY}
\affil[1]{Sorbonne Universit\'{e}, IRD, UMMISCO, F-93143, Bondy, France} 
\affil[2]{Universit\'{e} de Yaound\'{e} I, CETIC, FS, D\'{e}partement d'Informatique, BP 812, Yaound\'{e}, Cameroun} 
\affil[3]{Sorbonne Universit\'{e}, CNRS, Laboratoire d'Informatique de Paris 6, LIP6, F-75005, Paris, France} 
\corrauthor{armel.nzekon@lip6.fr}

\doi{10.18713/JIMIS-ddmmyy-v-a}{}
\review{February 27, 2019}{Month-in-letters Day Year}
\publication{5}{2019}

\issue{Analyse de graphes et r\'{e}seaux}
\editors{Didier Josselin, Vincent Labatut}

\begin{document}

\maketitle

\abstract{Recommending appropriate items to users is crucial in many e-commerce platforms that contain implicit data as users' browsing, purchasing and streaming history. One common approach consists in selecting the N most relevant items to each user, for a given N, which is called top-N recommendation. To do so, recommender systems rely on various kinds of information, like item and user features, past interest of users for items, browsing history and trust between users. However, they often use only one or two such pieces of information, which limits their performance. In this paper, we design and implement GraFC2T2, a general graph-based framework to easily combine and compare various kinds of side information for top-N recommendation. It encodes content-based features, temporal and trust information into a complex graph, and uses personalized PageRank on this graph to perform recommendation. We conduct experiments on Epinions and Ciao datasets, and compare obtained performances using F1-score, Hit ratio and MAP evaluation metrics, to systems based on matrix factorization and deep learning. This shows that our framework is convenient for such explorations, and that combining different kinds of information indeed improves recommendation in general.}


\section{Introduction}

Many e-commerce platforms have large and fast growing sets of items to present to users. For instance, Amazon had a total of 53.38 millions books as on January 10th, 2018\footnote{https://www.scrapehero.com/many-products-amazon-sell-january-2018/}. Such huge quantities of products make it challenging for users to search and find interesting items for them. Then, they often rely on the help provided by recommender systems.

Various approaches co-exist, the most classical ones being rating prediction and top-N recommendation \cite{steck2013evaluation}. Rating prediction estimates the rating value that a user is likely to give to items. Top-N recommendation ranks items for a given user and selects the N most interesting ones, for a given N. Many research works are dedicated to rating prediction. This requires explicit rating data whereas, in many platforms dedicated for instance to e-commerce, ratings are not available, and recommender systems have to deal with implicit data such as users' purchase, browsing and streaming history. In such situations, top-N recommendation can still be carried out \cite{cremonesi2010performance}.

In addition to the previous remark, top-N recommender systems are everywhere from on-line shopping websites to video portals \cite{christakopoulou2016local}. For all these reasons, we focus here on top-N recommendation problem from positive implicit feedback, a problem already considered in many papers such as \cite{rendle2009bpr, ning2011slim, shi2012climf} and \cite{guo2017factored}.

One of the main families of techniques, called Collaborative Filtering (CF), takes benefit from correlations between user interests. Initially, CF recommender systems focused only on user-item interactions \cite{konstan1997grouplens, herlocker1999algorithmic, sarwar2001item} and did not integrate side information among the following list: item features like the genre of a movie or the author of a song, context of interactions like location, timestamps or weather, and trust between users. Since such side information strongly influences user choices (for instance, users may listen to a new song because they like the singer), performances of such systems may be limited. In addition, side information helps solving problems like cold start and data sparsity \cite{burke2002hybrid, adomavicius2005toward, massa2007trust, campos2014time}.

For these reasons, much effort was devoted to the inclusion of side information into CF techniques. For instance, hybrid systems incorporate item features in order to combine CF and content-based filtering (CBF)~\cite{burke2002hybrid, chen2016content, shu2018content}. Likewise, a winning team of the Netflix competition \cite{koren2009matrix, koren2010collaborative} included temporal information into a CF system in order to track the dynamics of user interests and increase recommendation accuracy. Including trust information in order to take into account the fact that people tend to adopt items already chosen by trusted friends is also possible~\cite{papagelis2005alleviating,massa2007trust,guo2017factored}.

Some previous works consider only one type of side information, and therefore fail to capture the combined influence of several types of side information on user interests. Others works suggest that progress in this direction may significantly improve recommendation, and combine two kinds of side information into CF \cite{ning2012sparse,yu2014topic,strub2016hybrid,nzeko2017time}. However, to the best of our knowledge, none of these approaches include content-based features, users' preferences temporal dynamics and trust relationships between users simultaneously. 

Our goal in this paper is to propose a general graph-based recommender framework that makes it easy to combine variety of side information. However, recommender systems are used in very diverse situations, which makes the design of a fully general system out of reach. We therefore made several assumptions which, although very general, do not apply to some contexts. First, we focus on top-N recommendation task because it is prevalent in many on-line shopping recommender systems like video portals. In addition, we considered the situations where the recommender system aims at offering each user a product that he/she has not yet selected in the past. In some situations, clients may repeatedly buy the same product, but this is a quite different problem. We also we assumed that recent activities are more important than older ones, a situation known as concept drift. This is often but not always true in practice; interest in a given kind of product may for instance be periodic, like for birthday gifts or seasonal needs. Extending our work in this direction is promising, when data is available. Finally, we consider positive links only (that typically represent a purchase), as this is the most prevalent case in practice; considering more subtle feedback from users, and in particular negative feedback, is a very promising direction for future work.

\subsection*{Contribution}

In this paper, we propose GraFC2T2, a general graph-based framework for top-N recommendation combining content-based features, temporal information, and trust into a personalized PageRank system. The design of this framework is very modular in order to make it easy to include other side information and/or replace personalized PageRank by another graph-based method. Thanks to GraFC2T2, it becomes easy to explore the benefit of using various kinds of side information, and then to find appropriate parameters for combining them for particular applications. We conduct experiments on Epinions and Ciao datasets to illustrate the use of GraFC2T2, and we show that it outperforms state-of-the-art thanks to the increased use of side information.

\begin{figure}[!h!]
\includegraphics[height=0.7\textwidth,width=1\textwidth]{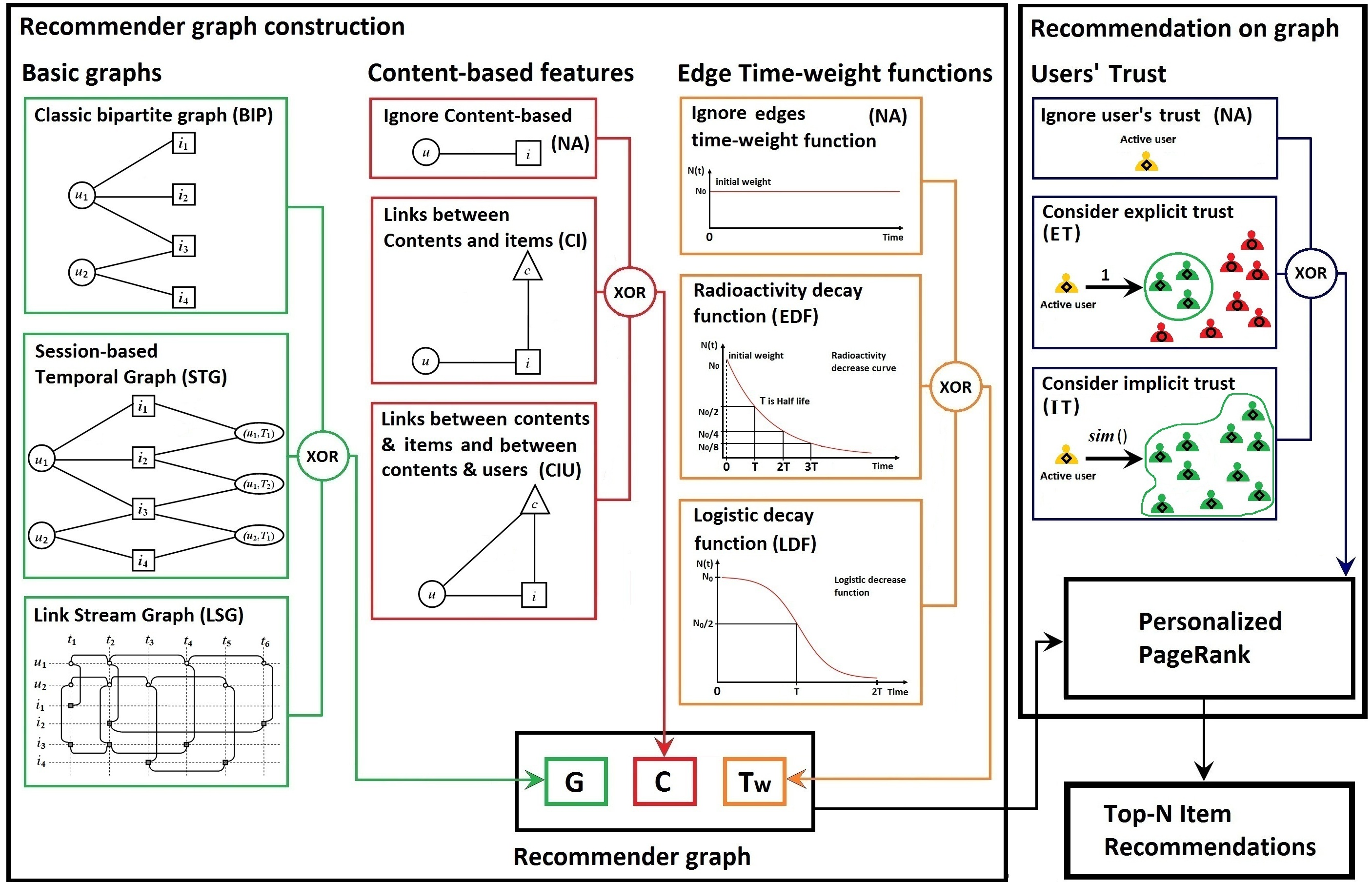}
\caption{{\bf The global architecture of GraFC2T2}, our general purpose graph-based recommender framework. Recommender graphs are built from three components: a basic graph that models user-item relations, content-based features that enrich basic graph, and link time-weight function that penalizes old edges, see Sections~\ref{sec:basic_rs_graph} and~\ref{sec:graph_extension}. Then, we perform top-N recommendation over this graph using user trust and personalized PageRank, see Section~\ref{sec:pagerank}.}
\label{fig_grafctt_architecture}
\end{figure}

Figure~\ref{fig_grafctt_architecture} summarizes the global architecture of GraFC2T2, made of two big parts: the recommender graph construction, and the use of this graph to perform recommendation. The recommender graph encodes available information by combining a basic graph, which we detail in Section~\ref{sec:basic_rs_graph}, with methods to capture content-based features and edge weight capturing time information, which we detail in Section~\ref{sec:graph_extension}. Then, we use the obtained recommender graph to perform recommendation, with a trust-aware personalized PageRank detailed in Section~\ref{sec:pagerank}.

Notice that our framework makes it possible to explore wide sets of modeling choices, as well as to incorporate additional possibilities if needed. We illustrate this on two real-world datasets from Epinions and Ciao in Sections~\ref{sec:experiments} and~\ref{sec:experiments_results}. Section~\ref{sec:relatedwork} discusses related work.

This work builds upon our previous paper \cite{nzeko2017time}, which extends the Session-based Temporal Graph proposed by \cite{xiang2010temporal} by adding time-weight and content-based information. On the other hand, the data representation that we use is the link stream formalism, presented in \cite{latapy2017stream}. This model allowed us to propose the Link Stream Graph \cite{nzekon2019link}.

We provide an implementation of our framework at \url{https://github.com/nzekonarmel/GraFC2T2} in order to help other researchers and practitionners to conduct experiments on their own datasets, and to test the relevance of new ideas and features.

\section{Data modeling}
\label{sec:basic_rs_graph}

\begin{figure} [!h]
\centering
\includegraphics[height=0.35\textwidth,width=0.5\textwidth]{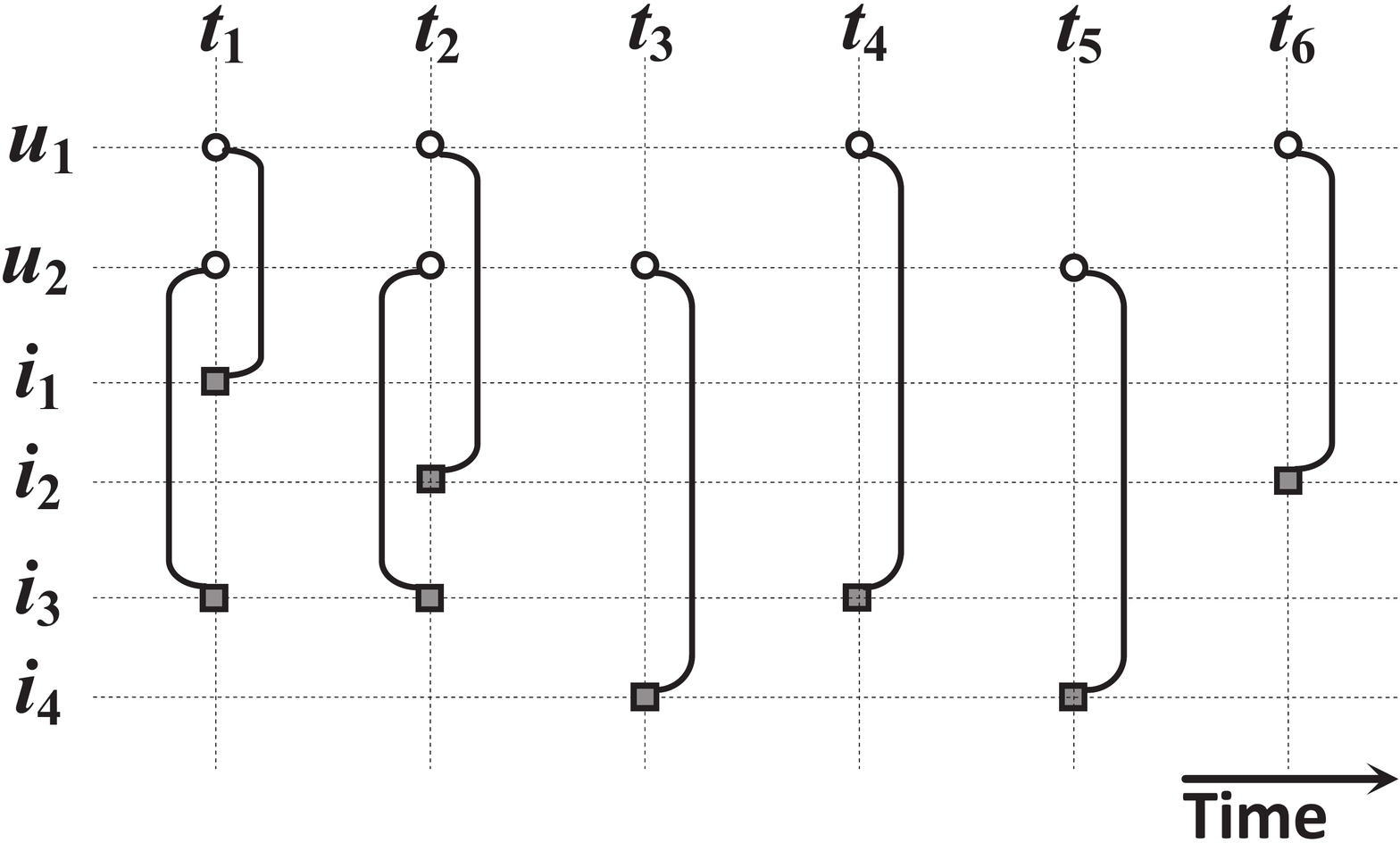}
\caption{{\bf Guiding example}: we consider the link stream $L=(T,U,I,E)$ in which the set of users is $U=\{u_1, u_2\}$, the set of items is $I=\{i_1,i_2,i_3,i_4\}$, the observation period is $T=[t_1, t_6]$, and $E = \{(t_1,u_1,i_1),\allowbreak (t_1,u_2,i_3),\allowbreak (t_2,u_1,i_2),\allowbreak (t_2,u_2,i_3),\allowbreak (t_3,u_2,i_4),\allowbreak (t_4,u_1,i_3),\allowbreak (t_5,u_2,i_4),\allowbreak (t_6,u_1,i_2)\}$. This means for instance that user $u_1$ was interested in item $i_2$ at time $t_2$.}
\label{fig_linkstream_L}
\end{figure}

We consider a set $U$ of users, a set $I$ of items, and a time interval $T$, and we assume that we observed the past interest of users in $U$ for items in $I$ during $T$. We model this data by a bipartite link stream $L=(T,U,I,E)$ where $E\subseteq T\times U\times I$ is a set of links: each link $(t,u,i)$ in $E$ represents a purchase ($u$ bought product $i$ at time $t$), an interest in a cultural item (like movie watching or song listening), or another user-item relational event, depending on the application context. See \cite{viard2016computing,latapy2018stream} for a full description of the link stream formalism. In the following, we will illustrate definitions with the guiding example of Figure~\ref{fig_linkstream_L}.

\subsection{Classical bipartite graph}

We first consider the most classical recommender graph introduced in the literature \cite{huang2004applying,baluja2008video}, that we denote by BIP. It is a directed bipartite graph $(U, I, E')$ where $U$ and $I$ are the set of users and items defined above, and $E' \subseteq U\times I$ is the set of links defined by $E' = \{(u,i): \exists t\in T, (t,u,i)\in E\}$. In other words, $u$ is linked to $i$ in BIP if user $u$ was interested in item $i$ during the observation period. Figure~\ref{fig:basic_graphs}(a) displays the BIP graph for the guiding example.

\subsection{Session-based temporal graph}

In a first attempt to capture time information, we then consider Session-based Temporal Graphs proposed by \cite{xiang2010temporal}, that we denote by STG.

This graph encodes time information using a set $S$ of session nodes defined as follows. First, for a given $\Delta$, the observation interval $T$ is divided into $\frac{|T|}{\Delta}$ time slices $T_k=[(k-1)\cdot\Delta, k\cdot\Delta]$ of equal duration $\Delta$. Then, $S$ contains the couples $(u,T_k)$ such that there exists a link $(t,u,i)$ in $E$ with $t\in T_k$. In other words, each user leads to a session node $(u,T_k)$ in $S$ for each time interval $T_k$ during which this user was active.

This finally leads to the definition of STG as a tripartite graph $(U,I,S,E'')$ with $U$, $I$, and $S$ defined above, and $E'' = E' \cup \{((u,T_k),i): \exists t\in T_k, (t,u,i)\in E\}$. In other words, we add to BIP the nodes in $S$, and a link between each session node $(u,T_k)$ and the items selected by user $u$ during time slice $T_k$.
Figure~\ref{fig:basic_graphs}(b) shows the STG representation for the guiding example.

Notice that in the original model \cite{xiang2010temporal}, any link from $u$ to $i$ has a weight 1 and any link from $i$ to $u$ has a weight $\eta$, where $\eta$ is a parameter. For simplicity, we do not consider this parameter here (or, equivalently, $\eta=1$), but it may easily be added if needed.

\subsection{Link stream graph}

In order to capture time information while avoiding the drawbacks of choosing a time window size $\Delta$ like for STG, we introduce the following link stream graph, that we denote by LSG \cite{nzekon2019link}. 

This graph is first defined by a set of nodes representing users and items over time: $\{(t,u): \exists i, (t,u,i)\in E\} \ \cup \ \{(t,i): \exists u, (t,u,i)\in E\}$.
In other words, each user $u$ is represented by the nodes $(t,u)$ such that a link involves $u$ in $L$ a time $t$, and each item is represented similarly.

\begin{figure}[!h]
    \centering
    \begin{tabular}{c}
    \begin{tabular}{r r}
    	\begin{tabular}{@{}c@{}}
	        \includegraphics[height=0.2\textheight,width=0.22\textwidth]{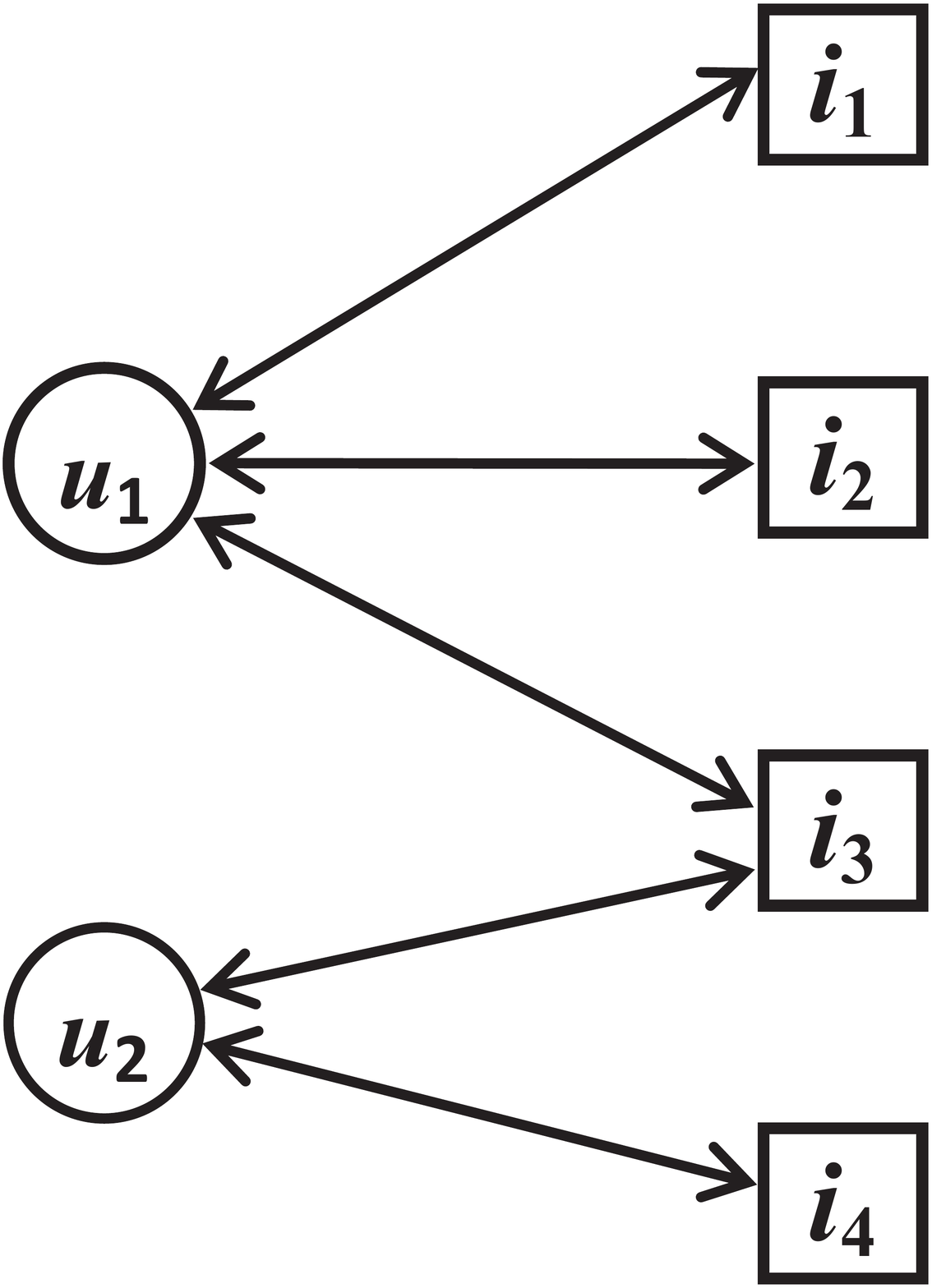} \vspace{3mm} \\
	        \small (a) Classical bipartite\\
	        \small graph, BIP
		\end{tabular}
		\hspace{5mm}
		&
		\hspace{5mm}
		\begin{tabular}{@{}c@{}}
		   \includegraphics[height=0.2\textheight,width=0.45\textwidth]{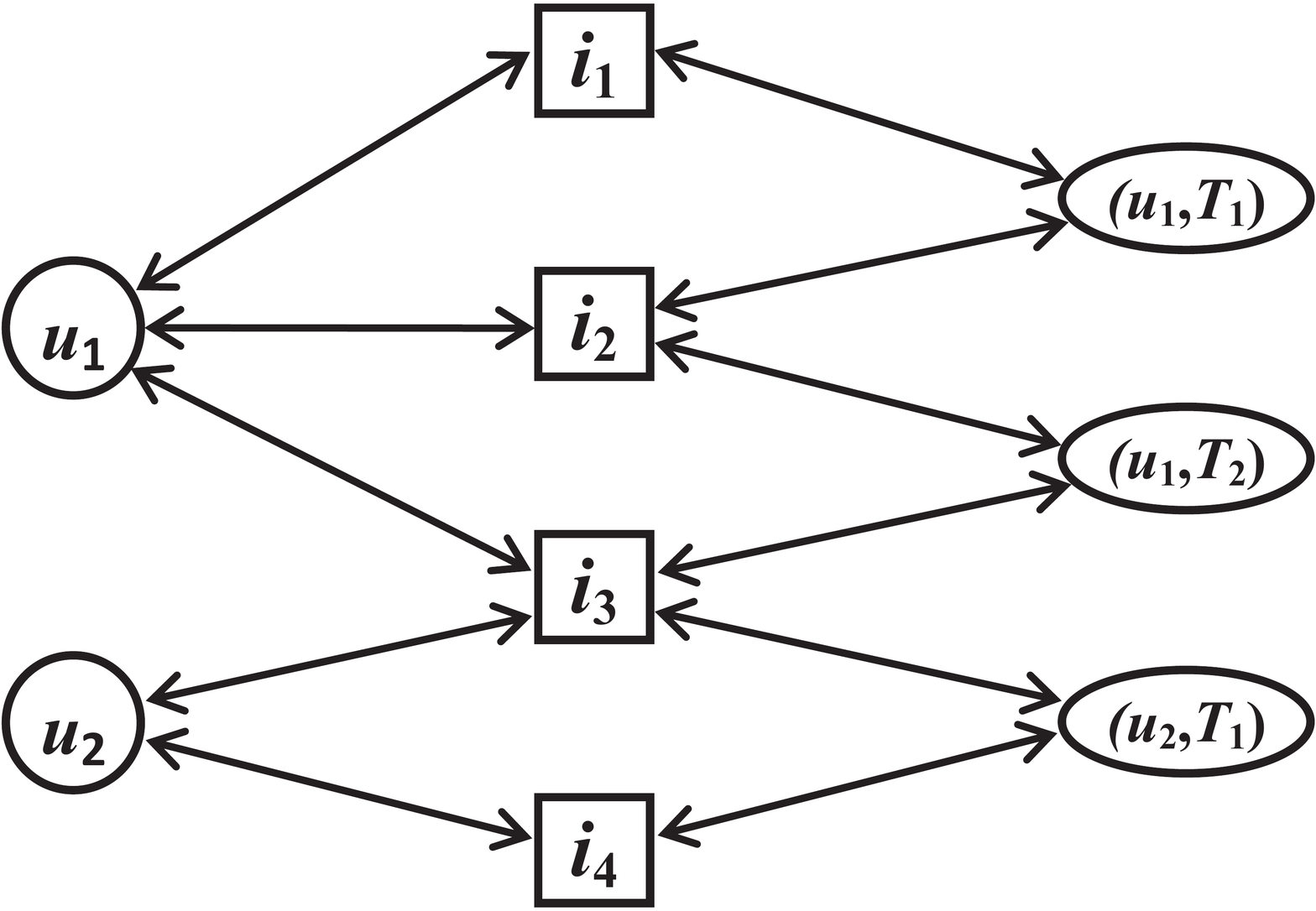} \vspace{3mm}\\
		   \small (b) Session-based temporal\\ 
		   \small graph, STG
		\end{tabular}	
	\end{tabular}
	\\
	\begin{tabular}{@{}c@{}}
		\includegraphics[height=0.25\textheight,width=0.65\textwidth]{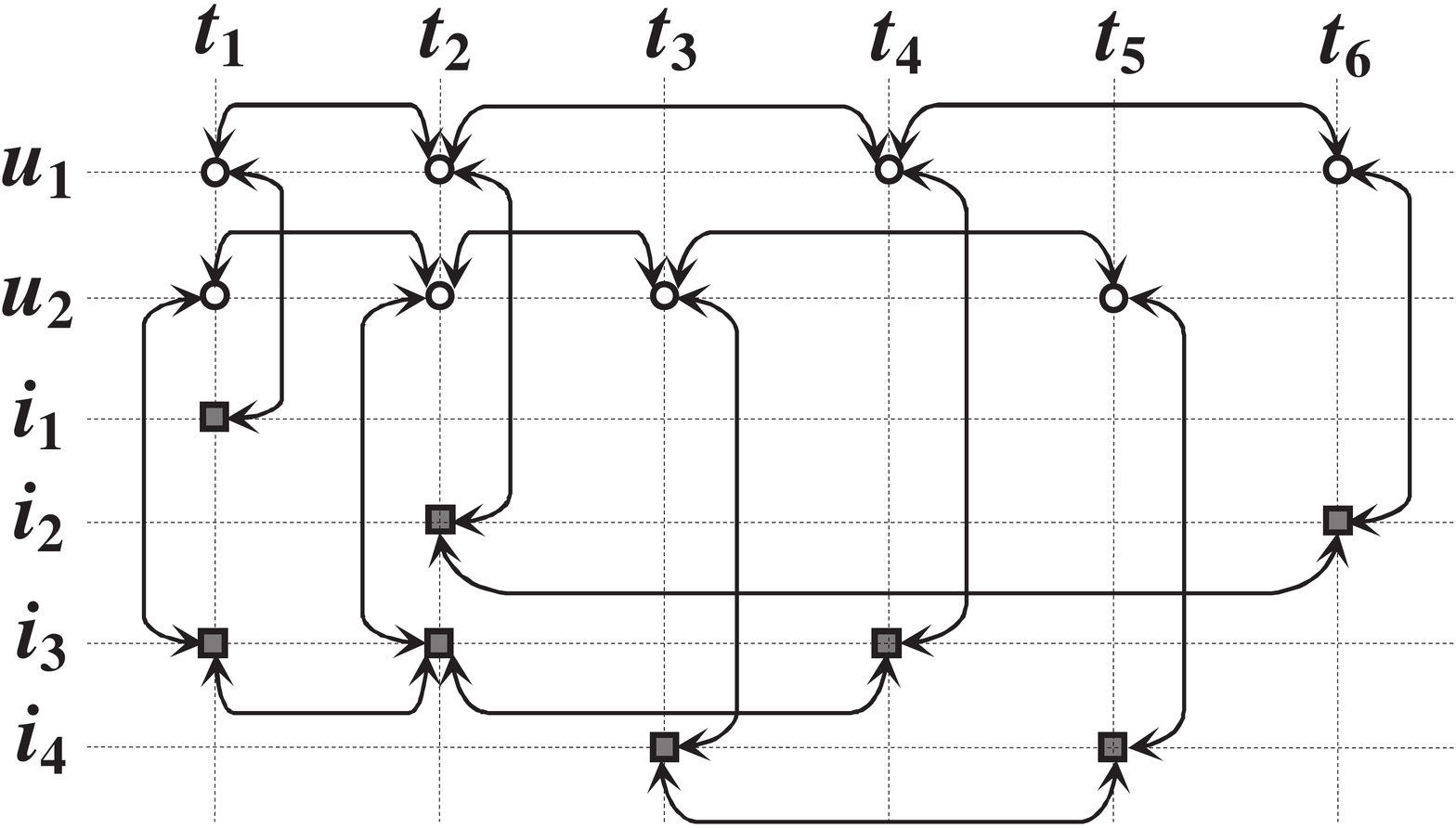} \\
		\small (c) Link stream graph, LSG 
	\end{tabular}
    \end{tabular}
    \caption{Classical bipartite graph, Session-based temporal graph and Link stream graph obtained from our guiding example. The weight of each edge is 1.} 
    \label{fig:basic_graphs}
\end{figure}

We then define the set of links $
\{((t,u),(t,i)): (t,u,i)\in E\}
\ \cup\ 
\{((t,u),(t',u)): \exists i, (t,u,i)\in E, t'=\min\{x:x>t\mbox{ and }\exists i', (x,u,i')\in E\}
\ \cup\ 
\{((t,i),(t',i)): \exists u, (t,u,i)\in E, t'=\min\{x:x>t\mbox{ and }\exists u', (x,u',i)\in E\}
$.
In other words, each user node $(t,u)$ is linked to both the item nodes $(t,i)$ such that $(t,u,i) \in E$ and to the next user node representing $u$. Item nodes are linked similarly. 
See Figure~\ref{fig:basic_graphs}(c) for an illustration on our guiding example.

\section{Adding content-based features and time-weight functions}
\label{sec:graph_extension}

Once a basic recommender graph is built as explained in previous section, the GraFC2T2 framework adds elements to capture content-based and temporal features. Again, we propose several choices, and we present them below.
  
\subsection{Content-based features}

Let $C$ be the set of all possible content-based features and let $g(i) \subseteq C$ be the subset of content-based features associated with item $i$, for any $i$. One element of $g(i)$ can be the category, the brand or the color of item $i$. Following the method proposed in \cite{phuong2008graph,yu2014topic,nzeko2017time}, we model these features by content nodes that we link to item nodes in basic recommender graphs.

\begin{figure}[!h]
    \centering
    \begin{tabular}{l l c}
       	\begin{tabular}{@{}c@{}}
       		\\
	        \includegraphics[height=0.12\textheight,width=0.3\textwidth]{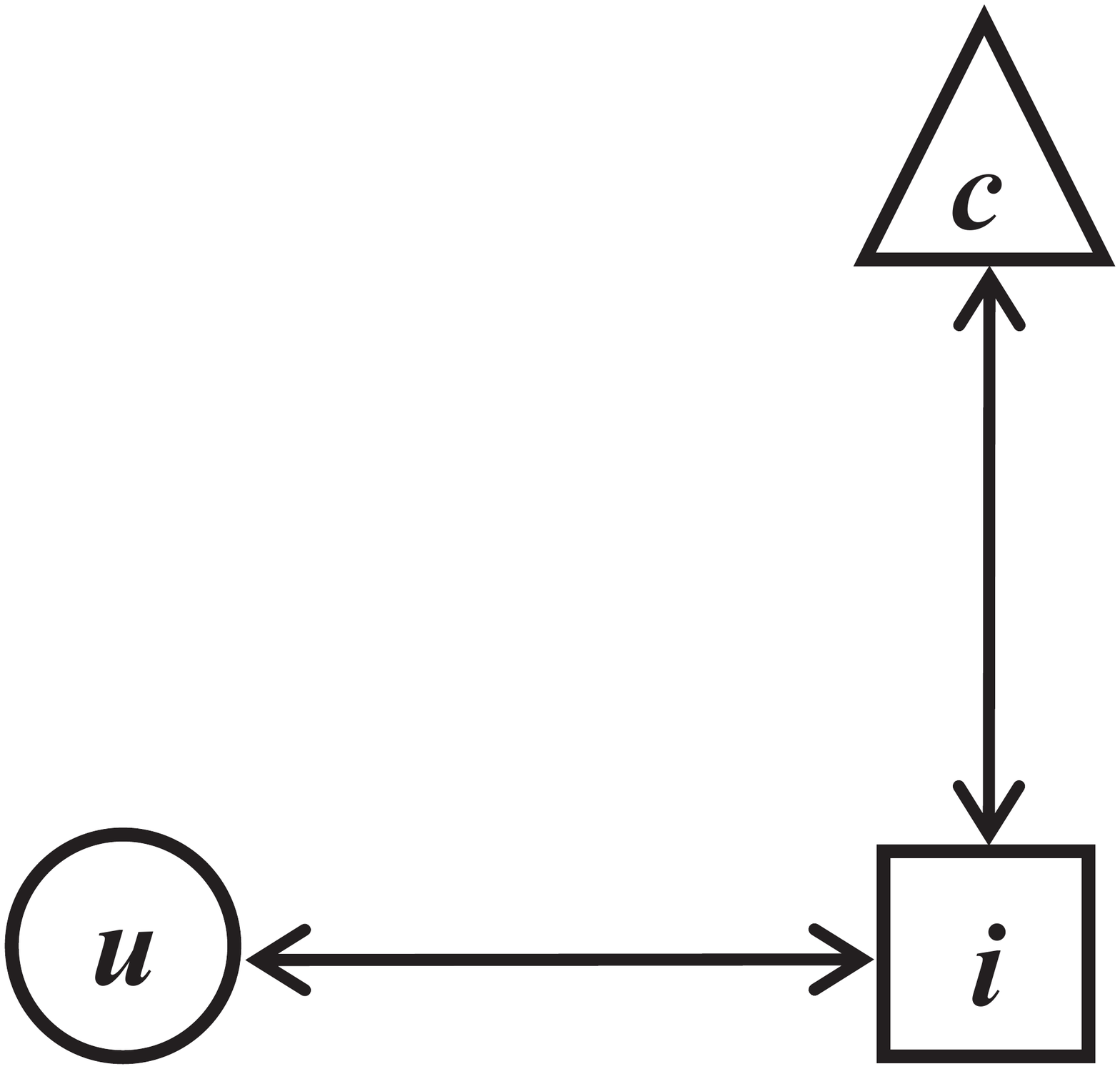} \\ \\
	        \small (a) BIP + CI
	        \label{fig:ci_bip}
   		\end{tabular}
   		&
   		\begin{tabular}{@{}c@{}}
		   \includegraphics[height=0.16\textheight,width=0.38\textwidth]{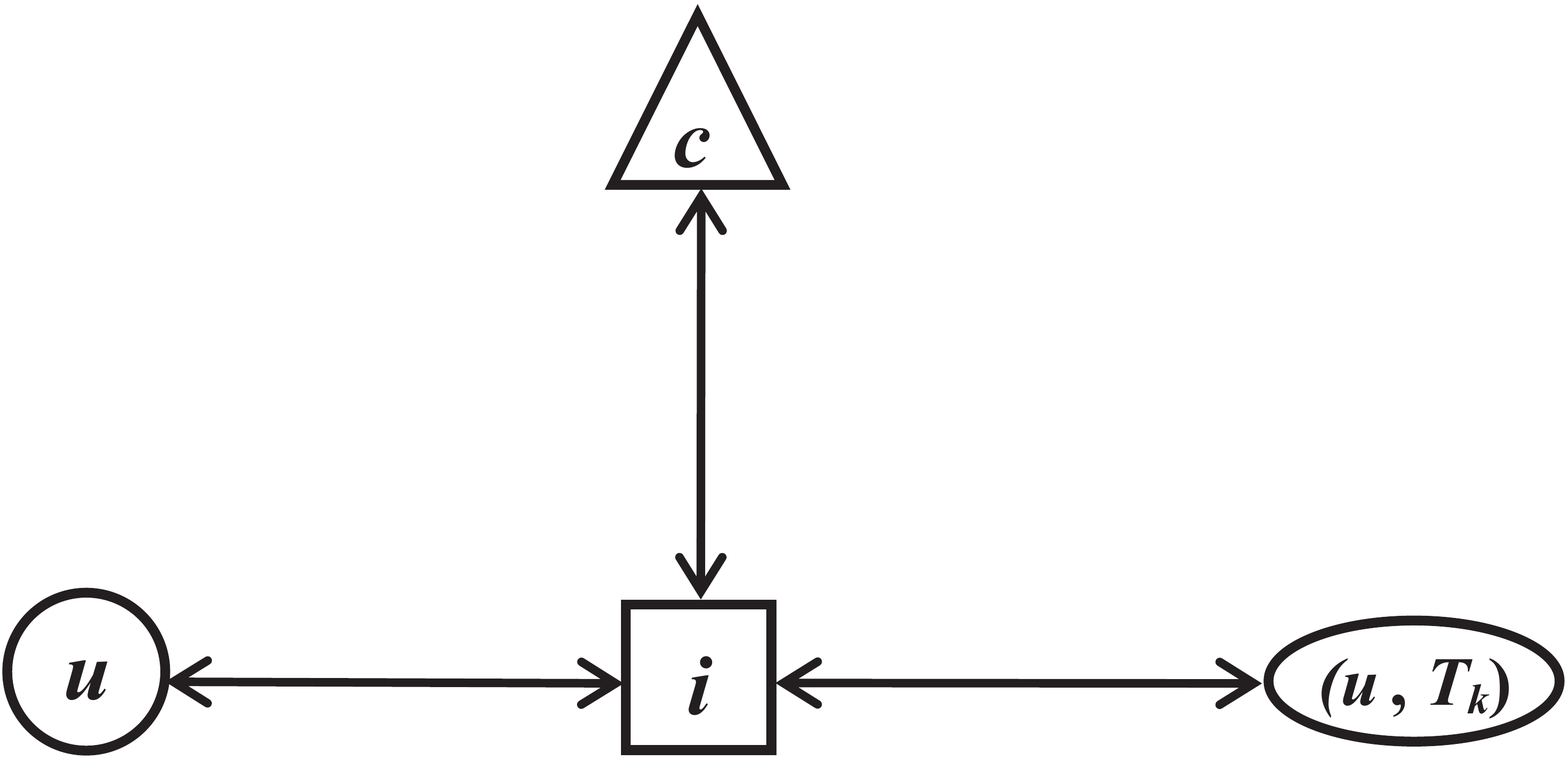} \\
		   \small (b) STG + CI
           \label{fig:ci_stg}
   		\end{tabular}
   		&
   		\begin{tabular}{@{}c@{}}
   			\includegraphics[height=0.16\textheight,width=0.2\textwidth]{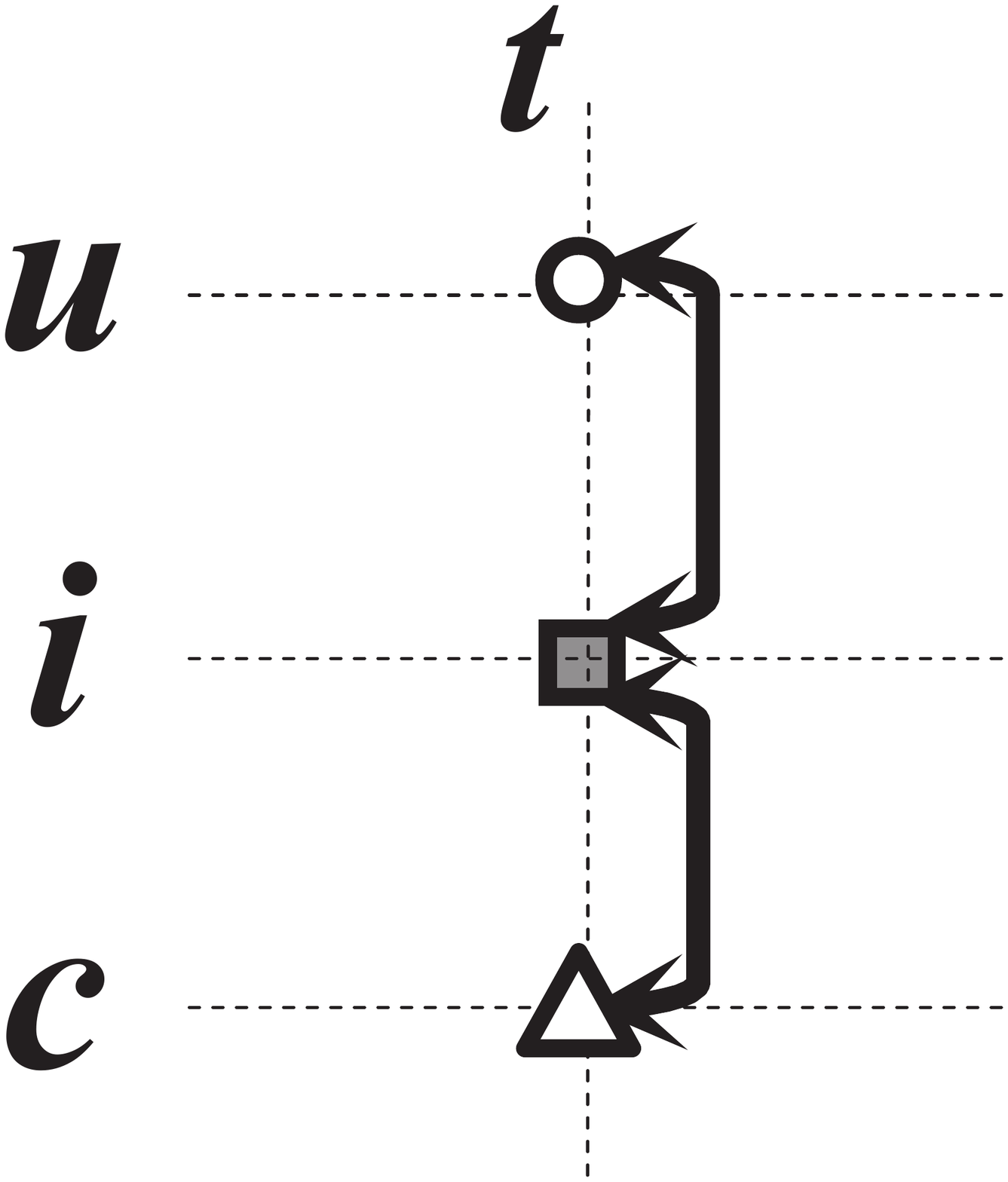} \\
   			\small (c) LSG + CI
   		    \label{fig:ci_lsg}
   		\end{tabular}
   	\end{tabular}
    
    \caption{Inclusion of nodes and links representing content-based features with the CI strategy, for each basic recommender graph.}
    \label{fig:ci_method}
\end{figure}

In the cases of BIP and STG, we add a content node $c$ for each content-based feature $c$ in $C$, and we link each item node $i$ to the content node $c$ for each $c$ in $g(i)$. For LSG, we add a content node $(t,c)$ for each $(t,i)$ in the basic graph such that $c$ is in $g(i)$, and we link $(t,c)$ to $(t,i)$. We call this inclusion of content-based features CI because it adds links only between content and item nodes. See Figure~\ref{fig:ci_method}.

\begin{figure}[!h]
    \centering
        \begin{tabular}{c c c}
           	\begin{tabular}{@{}c@{}}
           	    \\
    	        \includegraphics[height=0.12\textheight,width=0.3\textwidth]{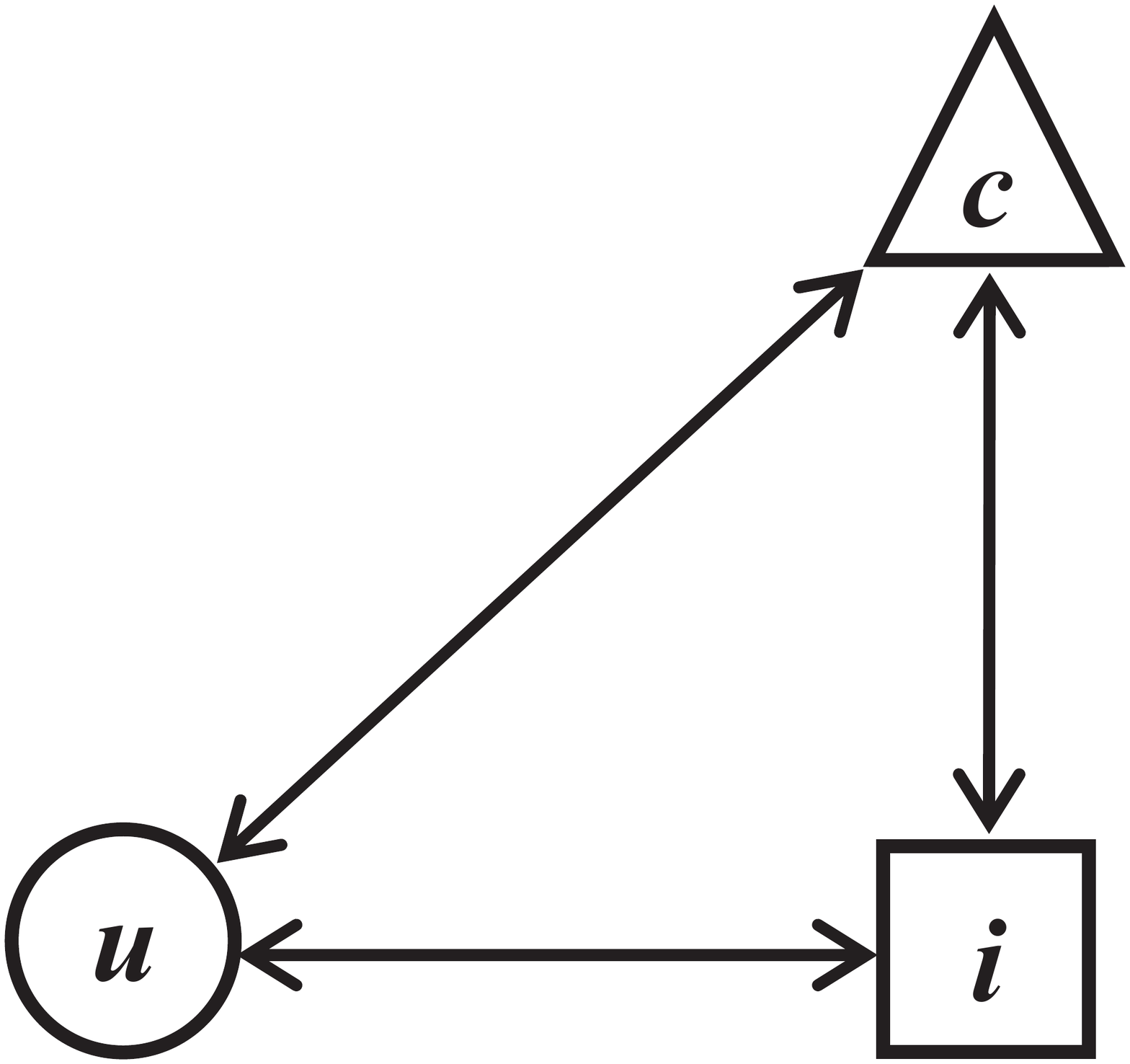} \\ \\
    	        \small (a) BIP + CIU 
    	        \label{fig:ciu_bip}
       		\end{tabular}
       		&
       		\begin{tabular}{@{}c@{}}
    		   \includegraphics[height=0.16\textheight,width=0.38\textwidth]{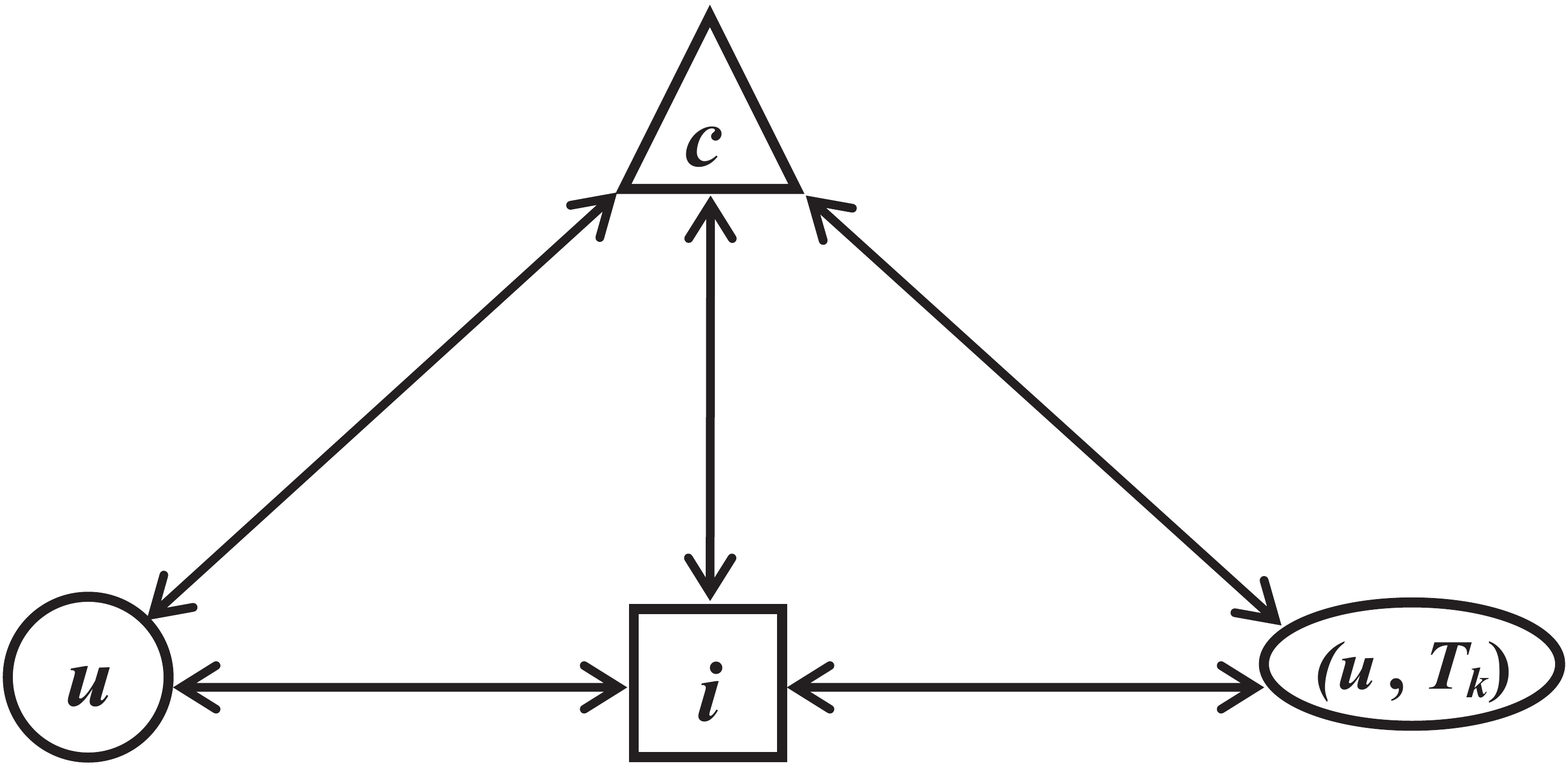} \\
    		   \small (b) STG + CIU
               \label{fig:ciu_stg}
       		\end{tabular}
       		&
       		\begin{tabular}{@{}c@{}}
       			\includegraphics[height=0.16\textheight,width=0.2\textwidth]{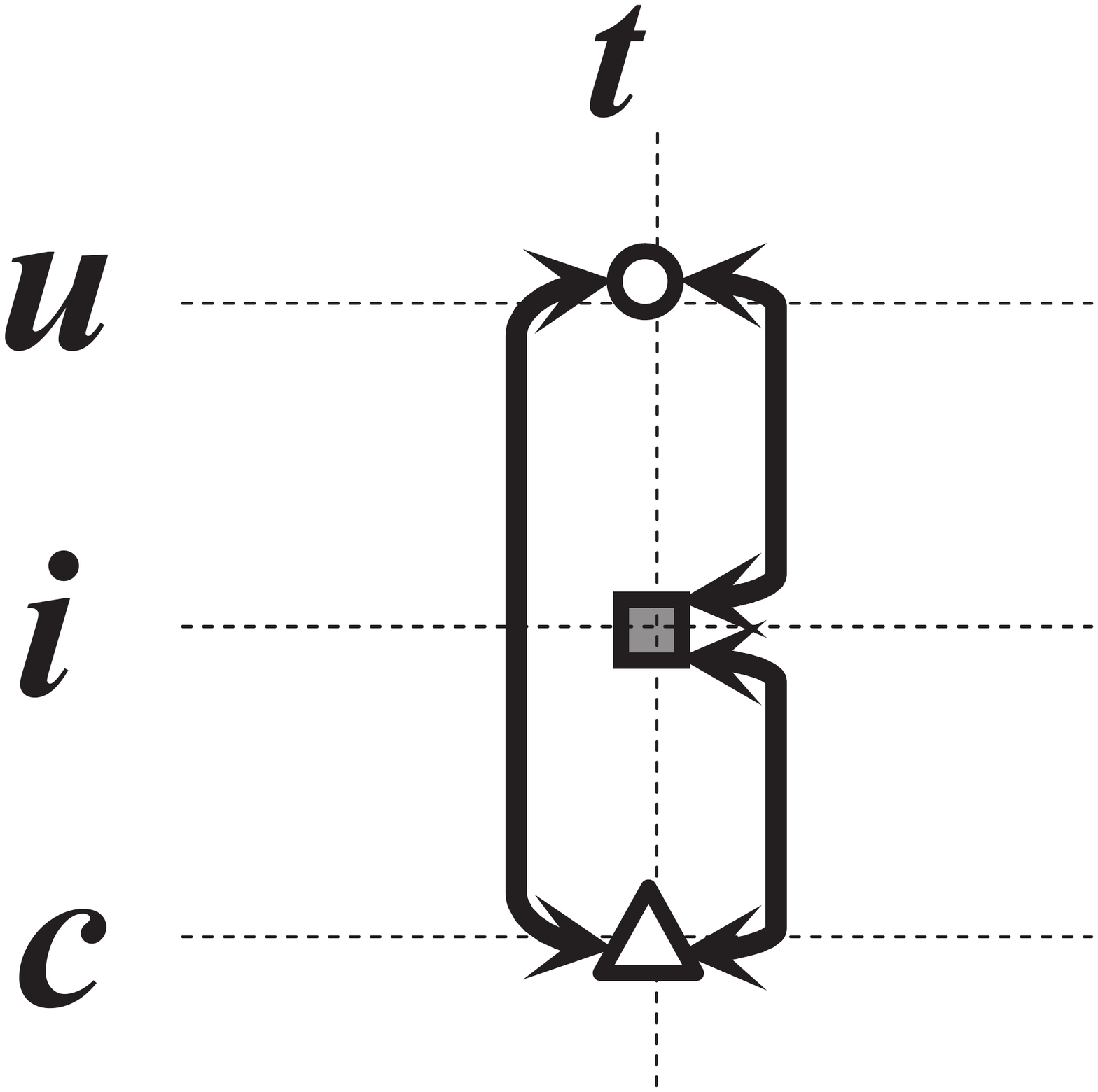} \\
       			\small (c) LSG + CIU
       		    \label{fig:ciu_lsg}
       		\end{tabular}
       	\end{tabular}
    
    \caption{Inclusion of nodes and links representing content-based features with the CIU strategy, for each basic recommender graph.}
    \label{fig:ciu_method}
\end{figure}

We also propose a strategy linking content nodes to both item and user nodes, that we call CIU. The idea is to link user nodes to the content nodes of the items they are interested in. Therefore, in addition to CI additions, CIU adds to BIP a link $(u, c)$ between each user node $u$ and content node $c$ whenever there is an item node linked to both $u$ and $c$; to STG a link between each session node $(u,T_k)$ and content node $c$ whenever there is an item node linked to both $(u, T_k)$ and $c$; and to LSG a link between each user node $(t,u)$ and content node $(t,c)$ whenever there is an item node linked to both. See Figure~\ref{fig:ciu_method}.

Compared to CI, the CIU method increases the influence of content-based features linked to items that the target user has already selected in the past. In other words, the CIU method do a better promotion of items that have the same features as the choices of the target user.

\subsection{Time-weight functions}

Until now, we modeled time information directly within the structure of STG and LSG graphs, but their edge weights give a static view of previous user interests. Since such interests evolve over time, as pointed out for instance in \cite{Ding2005}, this is not sufficient. We therefore follow the methodology proposed in that paper, consisting in adding time-dependent weights to the links of recommender graphs.

The idea is to give a high weight to recent links, and to decrease this weight with their age: the weight at time $t$ of any link $(a,b)$ whose most recent appearance time is $t_e\le t$, is of the form $w_t(a,b) = f(t - t_e)\cdot w(a,b)$, where $f()$ is a decay function. Many different decay functions may make sense, and we designed GraFC2T2 to make it easy to integrate those functions. We consider here the two following classical choices.

\begin{figure}[!h]
    \centering
        \begin{tabular}{c c}
           	\begin{tabular}{@{}c@{}}
    	        \includegraphics[height=0.18\textheight,width=0.45\textwidth]{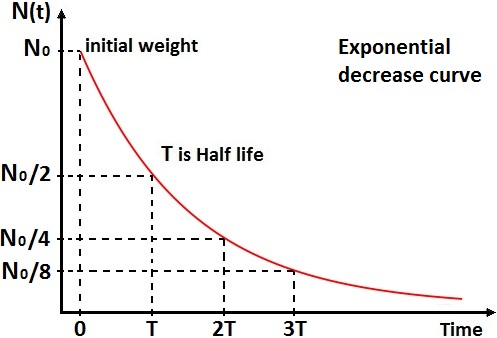} \\
    	        \small (a) Exponential decay function, EDF
    	        \label{fig:edf}
       		\end{tabular}
       		& 
       		\vspace{5mm}
       		\begin{tabular}{@{}c@{}}
    		   \includegraphics[height=0.18\textheight,width=0.45\textwidth]{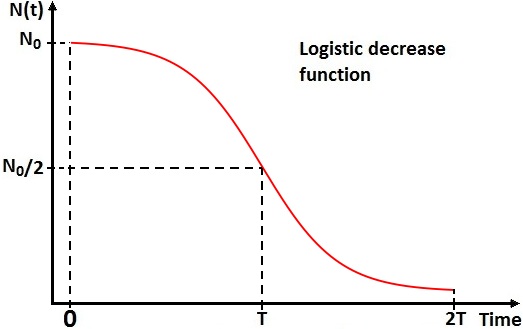} \\
    		   \small (b) Logistic decay function, LDF
               \label{fig:ldf}
       		\end{tabular}
       	\end{tabular}
    
    \caption{Edge time-weight functions.} 
    \label{fig:time}
\end{figure}

\begin{itemize}
\item Our first example is the exponential decay function (EDF) illustrated in Figure~\ref{fig:time}(a): $f(x) = e^{-x \cdot \ln(2) /\tau_0}$, where $\tau_0$ is the radioactivity half life; after a delay of $\tau_0$, the link weight is divided by $2$.

\item We also consider the logistic decay function (LDF) illustrated in Figure~\ref{fig:time}(b): $f(x) = 1 - 1/(e^{-K (x - \tau_0)} + 1)$ where $K$ is the steepness of the curve and $\tau_0$ is the sigmoid midpoint; if $x = \tau_0$ then $f(x) = 0.5$.
\end{itemize}

\section{Recommendation with Personalized PageRank and trust}
\label{sec:pagerank}  

Once a recommender graph is built with a combination of choices proposed in previous sections, we are ready to perform top-N recommendation from this graph. We present below the personalized PageRank approach and an extension to include the concept of trust between users.

\subsection{Personalized PageRank}

Personalized PageRank algorithm is defined by \cite{page1999pagerank} for node ranking in graphs so that nodes can be ranked efficiently in order of importance. The first application was on web pages, especially in the Google search engine. Then this algorithm has been widely used in recommender systems because of the good prediction quality obtained \cite{gori2007itemrank,kim2011personalized,csora2015pagerank}.

Following this last observation, \cite{xiang2010temporal} proposed the Temporal Personalized Random Walk (TPRW) to compute recommendations on STG. It was defined to tackle temporal recommendation using the personalization idea of \cite{haveliwala2002topic}, corresponding to the following formula:
\begin{equation}
\label{eq:pagerank}
PR = \alpha \cdot M \cdot PR + (1-\alpha) \cdot d
\end{equation}
Where $PR$ is PageRank vector that contains the importance of each node at the end of the propagation process that we want to compute; $M$ is the transition matrix of the considered graph; $\alpha$ is the damping factor; and $d$ is the personalization vector indicating which nodes the random walker will jump to after a restart. In other words, $d$ allows to initialize the weight of source nodes. This process favors the recommendation of products that are close to source nodes: items close to source nodes with large weights in vector $d$, are favored (see below).

For a given user $u$ at time $t$, we define the personalized temporal vector $d$ as follows, depending on the type of basic graph: 
\begin{itemize}
\item for BIP, the walker always restarts from $u$: $d(u) = 1$ and $d(v)=0$ if $v \ne u$;
\item for STG, the walker either restarts from $u$ or from its most recent session node $(u,T_k)$: $d(u) = \beta$, $d(u,T_k) = 1-\beta$, $d(v)=0$ if $v \ne u$ and $v \ne (u,T_k)$;
\item for LSG, the walker always restarts from the most recent temporal node representing $u$, $(t',u)$: $d(t',u) = 1$ and $d(t'',v)=0$ if $(t'',v) \ne (t',u)$.
\end{itemize}

Then, we run PageRank over the recommender graph to compute the interest of each user $u$ for item $i$ at time $t$, and output the $N$ items with highest interest (in LSG, the interest for item $i$ is the sum of interests for $(t,i)$, for all $t$).

\subsection{Trust integration}

Trust relationships are interesting for improving recommendation, especially for cold users and cold items (users or items for which very limited information is available). Some systems incorporate trust information explicitly specified by users~\cite{jamali2009using,guo2017factored,pan2017trust}, but since such explicit information is rarely available, several approaches infer implicit trust~\cite{pitsilis2004model,papagelis2005alleviating,hwang2007using,lathia2008trust}. In this section, we describe how to include these both types of trust in our framework.

We assume trust relationships are modeled for each user $u$ by a set $TR_u$ of users trusted by $u$, and that $trust(u,v)$ gives the trust level of $u$ for all $v$ in $TR_u$, with $\sum_{v\in TR_u} trust(u,v) = 1$. We denote the method where explicit trust relationships are given by ET (Explicit Trust). We also use an implicit trust metric based on similarity measures as proposed in~\cite{papagelis2005alleviating} and denote this method by IT (Implicit Trust). In this method, $TR_u = U$ is the set of all users, and $trust(u,v) = |I_{u} \cap I_{v}|/|I_{u} \cup I_{v}|$ is the Jaccard similarity between users $u$ and $v$. Note that other similarity measures may be used, such as cosine index.

We then update the personalized temporal vector $d$ definition as follows (with the same notations as in the initial definition above): 
\begin{itemize}
\item for BIP, $d(u) = 1 - \gamma$, $d(v) = (\gamma \cdot trust(u,v))/|TR_u|$ if $v \in TR_u$ and $d(v)=0$ otherwise;
\item for STG, we share the jumping probability  $\beta$ between $u$ and its trusted users:
$d(u) = \beta \cdot (1 - \gamma)$, $d(v) = (\beta \cdot \gamma \cdot trust(u,v))/|TR_u|$ for all $v\in TR_u$; and we share the probability $1- \beta$ between $u$ most recent session node and the ones of trusted users: $d(u,T_k) = (1 - \beta) \cdot (1 - \gamma)$, $d(v,T_v) = (1 - \beta) \cdot \gamma \cdot trust(u,v)/|TR_u|$ where $v \in TR_u$ and $(v,T_v)$ is the most recent session node of $v$. We set all other entries of $d$ to $0$.
\item for LSG, $d(t_k,u) = 1 - \gamma$, $d(t_v,v) = \gamma \cdot trust(u,v)/|TR_u|$ if $v \in TR_u$ and $(t_v,v)$ is the most recent node representing $v$, and all other entries of $d$ are 0.
\end{itemize}

\section{Experimental setup}
\label{sec:experiments}

Previous sections defined our general graph-based framework GraFC2T2, that gives wide levels of freedom for selecting and combining its various components into a top-N recommender system. These component capture several kinds of side information, in particular content-based, temporal, and trust features. In this section, we describe an experimental setup that we use in the next section to evaluate our framework. This setup consists in two real-world datasets, an evaluation method relying on three metrics, and a parameter selection method to optimize results.

\subsection{Datasets}

We use publicly available datasets extracted from product reviews Epinions and Ciao\,\footnote{\url{https://www.cse.msu.edu/~tangjili/trust.html}} \cite{tangetal12a}, where users can write reviews and give their opinions on a wide category of products like Home, Health, Computers and Media. We model each dataset as a set of review tuples $(u,i,c,r,t)$ meaning that user $u$ has assigned the rating $r \in \{0,1,2,3,4,5\}$ to item $i$ at time $t$, with $c$ being a content-based feature of item $i$. The explicit trust networks of these datasets are considered such that for each user $u$, the set $TR_u$ is given for the ET method. Table \ref{tab_dataset} provides key information on these datasets: start and end dates, as well as numbers of distinct users, items, content-based features, ratings, explicit trust relationships, ratings density and trust relationships density.

\begin{table}[h]
\caption{Basic data statistics}
\label{tab_dataset}
\centering
\setlength\tabcolsep{4.5pt}
\begin{tabular}{|l|c|c|c|c|c|c|c|c|c|c|}
	\hline
		& start date  & end date & $\|U\|$ & $\|I\|$ & $\|C\|$ & ratings & trust & $\delta_r$ & $\delta_t$ \\
	\hline
	 Epinions & 2010-01-01 & 2010-12-31 & 1\,843 & 15\,899 & 24 & 17\,722 & 4\,867 & 0.06\% & 0.14\% \\
	\hline
	 Ciao & 2007-01-01 & 2010-12-31 & 879 & 6\,005 & 6 & 8\,109 & 23\,121 & 0.15\% & 3\% \\
	\hline
\end{tabular}
\end{table}

Since our framework does not use ratings but only positive links between users and items, we discard all tuples such that the rating it contains is lower than $2.5$ or the average rating of involved user.

\subsection{Evaluation}

Evaluating recommender systems is a difficult task. In this paper, we use three classical metrics for top-N recommendations: F1-score (F1), Hit Ratio (HR) and Mean Average Precision (MAP)~\cite{baeza2011modern}. Higher values of these metrics indicate better recommendation performance.

F1-score is a trade-off between ranking precision and recall such that optimizing F1-score is more robust than optimizing precision or recall. Precision is the fraction of good recommendations over all recommended items and recall is the fraction of good recommendations over all relevant items to recommend. For one user $u$, $Precision = \frac{hit_N(u)}{N}$ , $Recall = \frac{hit_N(u)}{I_{new}(u)}$ and $F1 = 2 \cdot \frac{Precision \times Recall}{Precision + Recall} = 2 \cdot \frac{hit_N(u)}{I_{new}(u) + N}$ where $N$ is the length of recommendation list, $hit_N(u)$ denotes the number of good recommendations to $u$ in the top-N items and $I_{new}(u)$ is the set of new items to recommend to $u$. For all users the equation of F1-score is: $F@N = \frac{\sum_{u \in U} 2 \times hit_N(u)}{\sum_{u \in U} (I_{new}(u) + N)}$.

Hit Ratio is the fraction of users to whom the recommender system has made at least one good recommendation over all users: $H@N = \frac{\sum_{u \in U} (hit_N(u) > 0)}{|U|}$.

Mean Average Precision considers the order of items in the top-N recommendation in order to give better evaluation scores to results that recommend better items first: $M@N = \frac{\sum_{u \in U} AP_N(u)}{|U|}$ where $AP_N(u) =  \frac{1}{hit_N(u)} \sum_{k=1}^N \dfrac{hit_k(u)}{k} \times h(k)$ is the average precision of top-N recommendations done to user $u$ and $h(k) = 1$ if the $k$-th recommended item is a good recommendation and $0$ otherwise.

These metrics evaluate a given top-N recommendation. Since we actually can't perform recommendations on live users, we perform evaluation on past data described above.  
Following the classical method established by previous works \cite{li2008expertise,lathia2009temporal,campos2014time,nzeko2017time}, we partition data according to $k+1$ time windows of equal duration, and we use them as follow. For each of the $k$ first slices:
\begin{itemize}
\item we build recommender graphs that correspond to data of this slice and all previous slices (training set),
\item we compute top-N recommendations for users who have selected at least one new item in the next time slice (test set),
\item we compute for each evaluation metric $M$ the numerator $M_{num_k}$ and the denominator $M_{deno_k}$ of its definition, given above.
\end{itemize}

Once we have the values of $M_{num_k}$ and $M_{deno_k}$ of each of the $k$ first windows, we combine them into the Time Averaged (TA) value of the metric under concern:
$TA(M) = \frac{\sum_{k} M_{num_k}}{\sum_{k} M_{deno_k}}$. This leads to a time-averaged value of F1-score, Hit ratio and MAP, that we all use for evaluation. Indeed, evaluation metrics can be in disagreement~\cite{gunawardana2009survey}, and so using several metrics is essential to obtain accurate insight on result quality.

In our experiments, we set $k$ to $7$ in order to have large enough data slices and meaningful averages. We consider exploring the role of this parameter, as well as the use of more advanced evaluation metrics, as future work.

\subsection{Parameter estimation}

For each basic graph type, GraFC2T2 defines and implements 27 possible combinations of side information modelings, see Figure~\ref{fig_grafctt_architecture}. Our priority is to explore the behaviors and differences of all these variants, and so we did our best to keep the number of other parameters reasonable. Still, the different version of recommender systems encoded in GraFC2T2 call for several parameter selection.

Exhaustive search for the best values is out of reach, and many subtle techniques exist to explore the parameter space in search for good values. Since this search is not the focus of this paper, we use a simple approach called Randomized Search Cross-Validation (more advanced methods may easily be included in our framework, though)~\cite{bergstra2012random}. This method randomly selects parameter values in a predefined set of possible values, usually designed to span well the whole set of values. Here, we use 50 such random settings, sampled in the set defined by Table~\ref{tab_param}.

\begin{table}[!h]
\caption{Predefined values of parameters}
\label{tab_param}
\centering
\begin{tabular}{|c|l|l|}
	\hline
	& \textbf{parameter meaning} & \textbf{predefined values} \\
	\hline
	$\boldsymbol{\Delta}$
	& STG session duration 
	& 7, 30, 90, 180, 365, 540, 730 \small{days}\\
	\hline
	$\boldsymbol{\beta}$
	& STG long-term preference
	& 0.1, 0.3, 0.5, 0.7, 0.9 \\
	\hline
	$\boldsymbol{\tau_0}$ 
	& half life of EDF and LDF
	& 7, 30, 90, 180, 365, 540, 730 \small{days}\\
	\hline
	$\boldsymbol{K}$ 
	& decay slope of LDF 
	& 0.1, 0.5, 1, 5, 10, 50, 100\\ 
	\hline
	$\boldsymbol{\gamma}$ 
	& influence of trusted users
	& 0.05, 0.1, 0.15, 0.3, 0.5, 0.7, 0.9 \\
	\hline
	$\boldsymbol{\alpha}$ 
	& damping factor for PageRank
	& 0.05, 0.1, 0.15, 0.3, 0.5, 0.7, 0.9 \\
	\hline
\end{tabular}
\end{table}

\section{Experimental results}
\label{sec:experiments_results}

This section presents extensive experimentations on our GraFC2T2 framework, in order to study its performances in practice, to explore the contribution of each side information in these cases, and to compare obtained results to state-of-the-art recommender systems.

\subsection{Performances of GraFC2T2}

Table~\ref{tab_result_prl} presents the results we obtained for Top-10 item recommendation for Epinions and Ciao datasets. We chose $N=10$ as for instance in \cite{deshpande2004item}, \cite{xiang2010temporal} and \cite{bernardes2015social}, and other values we tested gave similar results as one may see in the appendix (Section \ref{sec:appendix}). In these tables, each column corresponds to a metric and a basic recommender graph, and each row corresponds to a combination of side information added to this recommender graph. Each cell contains the value of the evaluation metric for the recommender graph made of basic graph in column and side information in row. White color of cell corresponds to the best result and dark color indicates lower performance.

\begin{table} 
\centering
\caption{Epinions and Ciao - Performance with optimal settings. Each cell contains the value of an evaluation metric for the recommender graph made of basic graph in column and side information in row. White color of cell corresponds to the best result and dark color indicates lower performance}
\label{tab_result_prl}
\begin{tabular}{c}
\includegraphics[height=0.65\textwidth,width=0.7\textwidth]{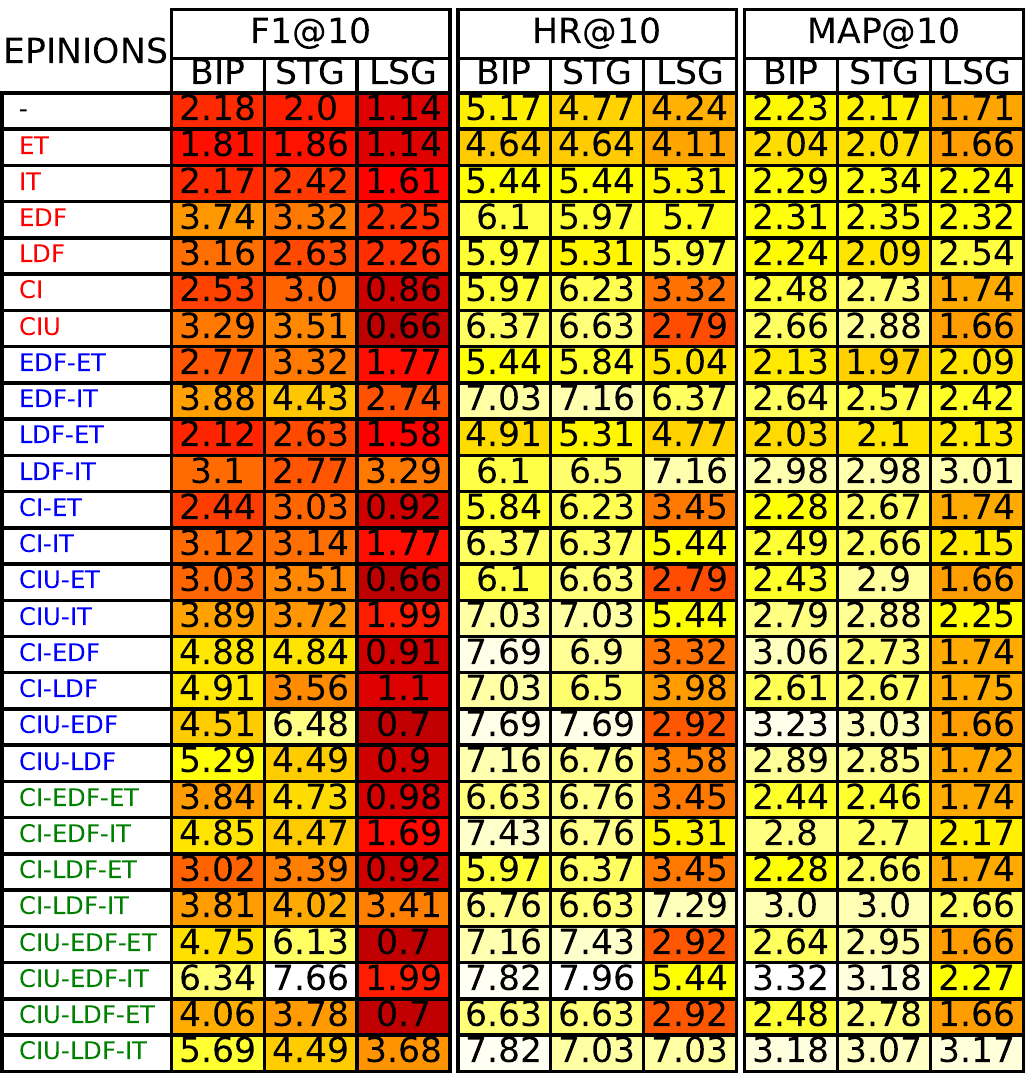}
\end{tabular}
\begin{tabular}{c}
\includegraphics[height=0.65\textwidth,width=0.7\textwidth]{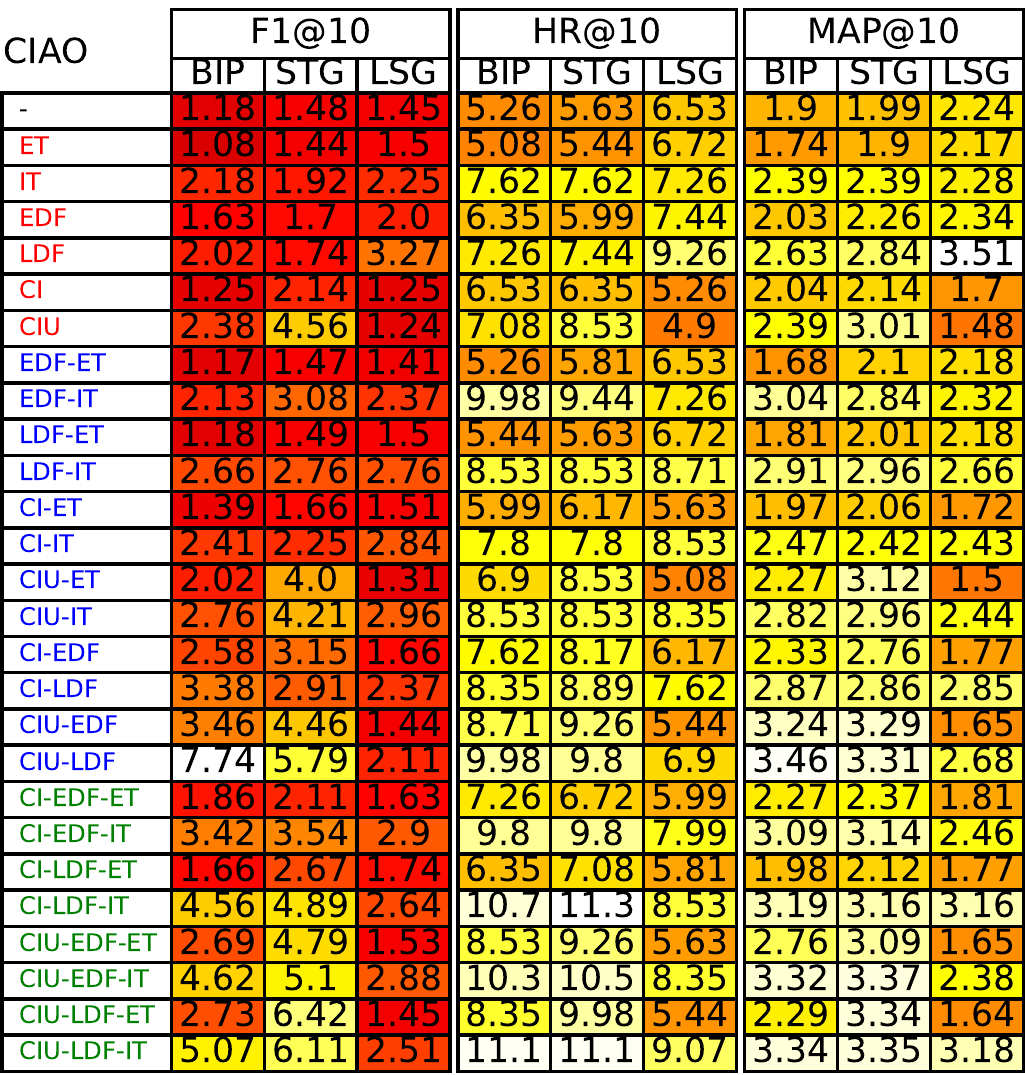}
\end{tabular}
\end{table}

We summarize the insight obtained from these results in Table~\ref{tab_imp_all}. For each basic recommender graph (vertically) and each evaluation metric (horizontally), we selected the three recommender graphs that achieve the best performances and we display on the corresponding row the performances obtained on the basic graph (without side information), the best obtained performances (with side information), the improvement percentage, and the name of the corresponding version of recommender graph with side information.

\begin{table} 
\centering
\caption{Best recommender graphs - Comparison of the three best recommender graph combinations with the associated basic graph. We display the obtained improvement percentage.}
\label{tab_imp_all}
\begin{tabular}{c}
\includegraphics[height=0.32\textwidth,width=0.95\textwidth]{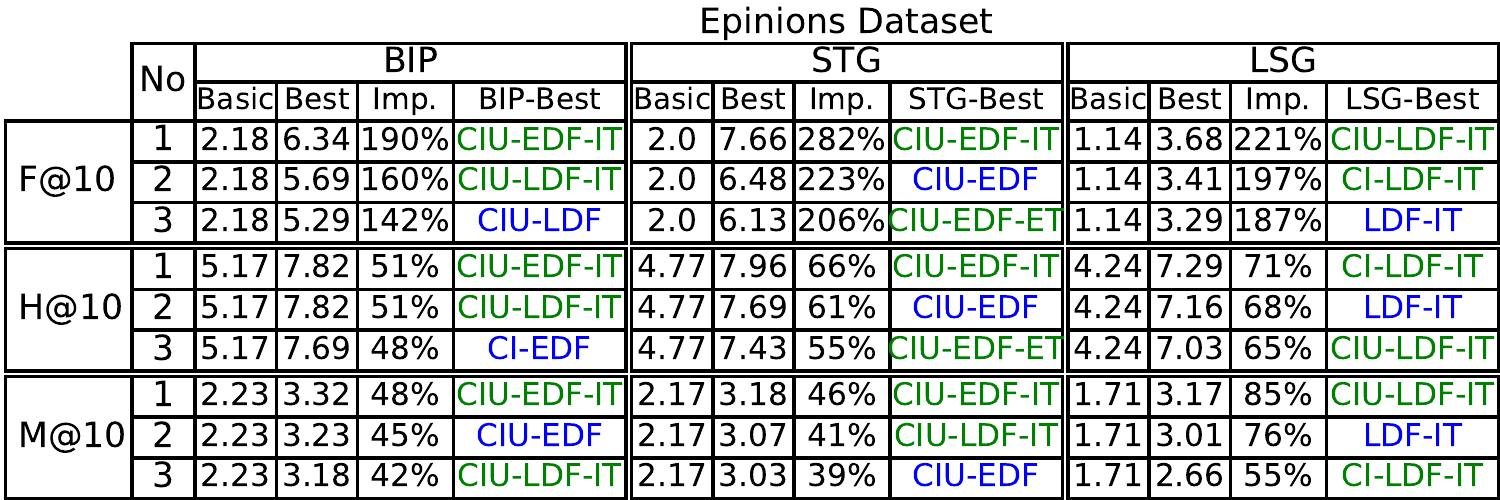}
\end{tabular}
\begin{tabular}{c}
\includegraphics[height=0.32\textwidth,width=0.95\textwidth]{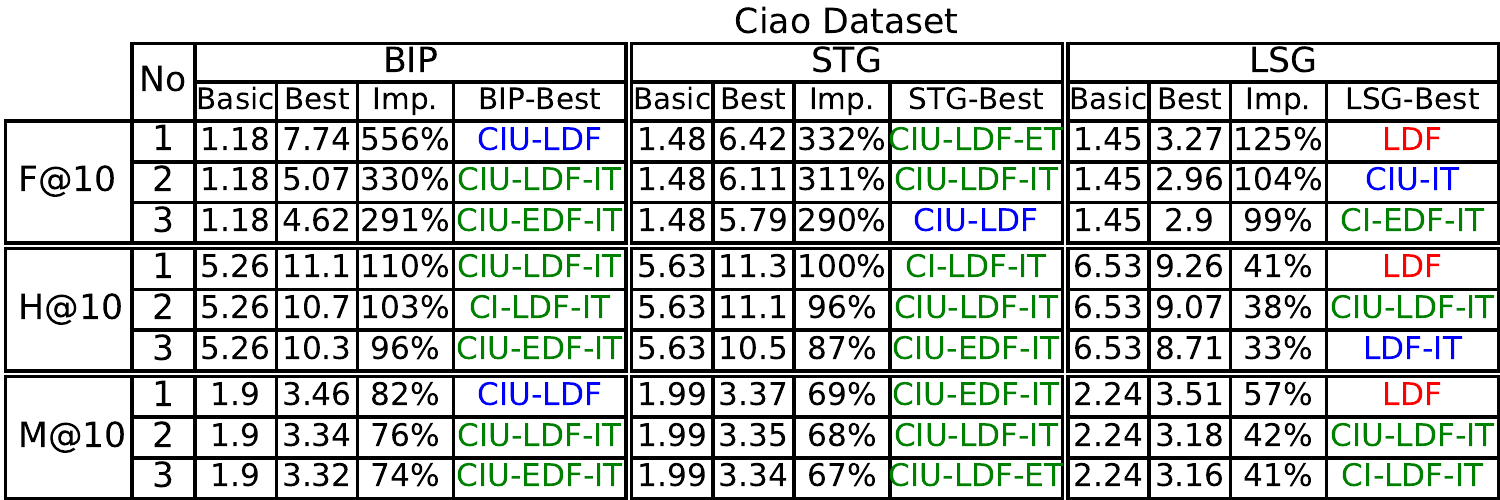}
\end{tabular}
\end{table}

All best improvements thanks to side information in GraFC2T2 are at least 46\% for Epinions and at least 41\% for Ciao. Table~\ref{tab_imp_all} also shows that the best combination of side information for Epinions is CIU-EDF-IT for BIP and STG basic graphs and CIU-LDF-IT for LSG basic graph. For Ciao, good results are obtained with CIU-LDF-IT for all basic graphs. These results clearly confirm the relevance of graphs extended simultaneously with content, time and trust information.

\subsection{Impact of side information}

We now give details on the impact of side information and their combination in GraFC2T2. This is context dependent, as observed behaviors vary with datasets; one may however easily test the GraFC2T2 framework with his/her own datasets and discover the best choices for the case under concern. The discussion provided here is mostly an illustration of this.

When we consider the basic graphs with no side information, in the case of Epinions, BIP gives the best results for all evaluation metrics. Instead, LSG gives the best Hit ratio and MAP, while STG gives the best F1-score in the case of Ciao. 

If we include only one kind of side information, we observe that explicit trust (ET) does not improve the results, but implicit trust (IT) does for all basic graphs. The insertion of time-weight always produces improvements. Finally, content-based features increase performances for BIP and STG but not for LSG. For Epinions, the best graph with one kind of side information is BIP-EDF in F1-score and STG-CIU in Hit ratio and MAP. In Ciao, the best one is LSG-LDF in Hit ratio and MAP, and STG-CIU is the best in F1-score. This shows that the impact of a unique kind of side information highly depends on the basic graph and on the data.

Recommendations using two kinds of side information perform significantly better than with only one kind of side information. For instance, in the Epinions case, performances increase from 3.74\% to 6.48\% in F1-score, from 6.63\% to 7.69\% in Hit ratio and from 2.88\% to 3.23\% in MAP. Combining time-weight with implicit trust performs better than time-weight and trust taken separately. Similarly, combining content-based features with implicit trust is better than content-based features or trust taken separately, but generally less interesting than combining time-weight and implicit trust. Combining content-based features and time-weight usually produces better improvements for BIP and STG but no improvement for LSG. In Epinions, BIP-CI-EDF and BIP-CIU-EDF perform best. In Ciao, BIP-CIU-LDF is always better. This confirms the relevance of graphs that integrate content-based features and time, like time-weight content-based STG proposed by \cite{nzeko2017time}.

Using three kinds of side information does not greatly improve the best performances achieved with two kinds of side information. For instance, in Epinions, the performances increase from 6.48 to 7.66\% in F1-score, from 7.69 to 7.96\% in Hit ratio and 3.23 to 3.32\% in MAP. Nevertheless, Table~\ref{tab_imp_all} shows that recommender graphs with three kinds of side information are by far the most frequent among the best ones. For this reason, we recommend the use of content-based, time and trust information simultaneously in order to increase the chances to achieve good results.

\subsection{Best values of parameters}

In this section, we focus only on recommender graphs with CIU-EDF-IT and CIU-LDF-IT combination that are most common in the best performance in Table~\ref{tab_imp_all}. We have made the following observations:

\begin{itemize}

\item In Epinions dataset, for the combination CIU-EDF-IT, $\Delta = 7$, $\beta = 0.5$, $\tau_0 = 90$ for BIP and STG and $180$ for LSG, $\gamma \in \{0.15, 0.3\}$ for BIP and STG and $0.9$ for LSG, and $\alpha = 0.9$. For the combination CIU-LDF-IT, $\Delta = 365$, $\beta = 0.7$, $\tau_0 \in \{30, 90\}$ for BIP and STG and $7$ for LSG, $K = 0.5$ for BIP, $100$ for STG and $5$ for LSG, $\gamma \in \{0.1, 0.15\}$ for BIP and STG and $0.9$ for LSG, and $\alpha \in \{0.7, 0.9\}$; 

\item In Ciao dataset, for the combination CIU-EDF-IT, $\Delta = 180$, $\beta = 0.3$, $\tau_0 = 180$, $\gamma = 0.9$ and $\alpha = 0.9$. For the combination CIU-LDF-IT, $\Delta = 540$, $\beta = 0.1$, $\tau_0 = 365$ for BIP and STG and $180$ for LSG, $K = 10$ for BIP and STG and $100$ for LSG, $\gamma \in \{0.7, 0.9\}$, and $\alpha = 0.9$; 

\end{itemize}

The values of these parameters indicate that in Epinions, the weights of the data used (edge weights) decrease faster than in Ciao; $\tau_0$ is small in Epinions $\{7, 30, 90\}$ and is larger in Ciao $\{180, 365\}$. Regarding trust, $\gamma$ is still high in Ciao $\{0.7, 0.9\}$ and is smaller in Epinions $\{0.1, 0.15, 0.3\}$ which shows that the influence of implicit trust is more important in Ciao. However, this influence must always be great for the graph LSG $\{0.9\}$ in all datasets.

\subsection{Comparison with state-of-the-art}

To evaluate our framework, we now compare its best performances with those of state-of-the-art Top-N recommender systems. The considered models are: the baseline system Most-Popular-Item (MPI) that computes the ranking score of an item by its popularity; the ranking oriented collaborative filtering, user-based (UBCF) and item-based (IBCF) collaborative filtering \cite{karypis2001evaluation,mclaughlin2004collaborative}; some state-of-the-art recommender systems for positive implicit feedback scenarios, Bayesian Personalized Ranking (BPR) \cite{rendle2009bpr}, Sparse linear methods for top-N recommender systems (SLIM) \cite{ning2011slim}, collaborative less-is-more filtering (CLiMF) \cite{shi2012climf} and Matrix factorization with Alternating Least Squares (ALS) \cite{hu2008collaborative}.

We use Randomized Search Cross-Validation to have good performances of the considered recommender systems. For UBCF and IBCF models, 10 settings are generated such that the neighborhood size $k \in \{10, 20, 30, 40, 50, 80, 100, 150, 200, 500\}$. For BPR, SLIM, CLIMF and ALS models, 50 settings are generated such that the number of latent factors $l \in \{10, 20, 30, 50,$ $100, 200, 500\}$, learning rate and all regularization bias are taken in $\{0.0001, 0.0005, 0.001,$ $0.005, 0.01, 0.05\}$. Table 5 presents the best results obtained for these recommender systems and comparison with those obtained with our framework. This shows that GraFC2T2 outperforms state-of-the-art recommender systems.

\begin{table}[h]
\caption{Experiment results on Epinions and Ciao datasets for Top-10. Performances are given in percentage and best ones are highlighted in bold.}
\label{tab_comparison}
\centering
\begin{tabular}{|l|l|c|c|c|c|c|c|c|c|c|}
\hline
\multicolumn{2}{|c|}{ }  
                 & MPI & UBCF & IBCF & BPR & SLIM & CLIMF & ALS & \small{GraFC2T2} \\
\hline
         & F@10  & 1.79 & 0.30 & 0.70 & 0.15 & 0.82 & 1.97 & 2.27 & \textbf{7.66} \\ \cline{2-10} 
Epinions & H@10  & 4.91 & 1.46 & 2.79 & 0.80 & 2.92 & 5.17 & 4.91 & \textbf{7.96} \\ \cline{2-10} 
         & M@10  & 2.07 & 0.61 & 1.29 & 0.45 & 1.16 & 2.15 & 2.26 & \textbf{3.32} \\ \hline
\hline
         & F@10  & 2.26 & 0.31 & 0.94 & 0.22 & 1.49 & 3.38 & 2.10 & \textbf{7.74} \\ \cline{2-10} 
Ciao     & H@10  & 7.62 & 1.63 & 4.17 & 1.27 & 5.08 & 8.71 & 6.90 & \textbf{11.3} \\ \cline{2-10} 
         & M@10  & 2.62 & 0.59 & 1.65 & 0.56 & 2.09 & 3.06 & 2.46 & \textbf{3.51} \\ \hline
\end{tabular}
\end{table}

Moreover, comparing the performance of our framework with results obtained for Epinions (MAP@10 = 1.32\%) and Ciao (MAP@10 = 3.07\%) by the Trust aware Denoising Auto Encoder (TDAE) technique based on deep learning~\cite{pan2017trust} confirms the relevance of GraFC2T2 framework. This shows that recommender graphs can reach performances comparable to and even better than matrix factorization and deep learning approaches, when the graphs are extended with content-based, temporal and trust information. 

Notice that the most basic, non-personalized approach MPI is able to achieve better results compared to BPR, SLIM, UBCF and IBCF. This indicates that users tend to consume popular items. This is not the first work in which MPI is better than BPR or other matrix factorization models, \cite{zhao2014leveraging} and \cite{guo2017factored} have made the same observation.

\section{Related work}
\label{sec:relatedwork}

As we already said, many contributions improve collaborative filtering (CF) recommender systems with the inclusion of side information, and we used several ideas proposed in these previous works. In the rest of this section, we shortly review key related references.

\subsection{Trust-based recommender systems}

CF usually suffers from data sparsity and cold start problems, which may be solved in part with user trust. For instance, \cite{papagelis2005alleviating} used trust inference by transitive associations between users in a social network. \cite{ma2017explicit} use explicit trust and distrust to improve clustering-based CF recommendation, while \cite{guo2014merging} merge ratings of trusted neighbors to infer probable preferences of other users, and identify similar users for item recommendations.

In some cases, trust can be explicitly provided by users as in \cite{massa2007trust}, but in other ones, this information is not given and it can be inferred from user behaviors. For example, in \cite{papagelis2005alleviating}, Pearson correlation is used to compute implicit trust using ratings dataset and in cases where there is only implicit data, measure like Jaccard and Cosine can be used. In other works, trust enhancement is done by trust propagation on trust network where the weight of an link $(u, u')$ is the trust of $u$ to $u'$ \cite{deng2014social}.

Note that work on influencers can also be considered here, as there is a trust relationship between influencers and their followers \cite{liu2015identifying,grafstrom2018impact}. Our framework is able to integrate the impact of influencers in the same way as trust between users. The main difference is who influences who and how much. Once you have the answers to these questions, the customization of PageRank is done according to these answers. The impact of influencers or influencer-based recommendation is not studied in this work, but it is a good issue for future work.

The concept of influence is a good example of other side information that may be included in our system \cite{liu2015identifying} and \cite{grafstrom2018impact}. Similarly to trust (although these two concepts are different) influence may be used to customize PageRank, once it is correctly quantified. For instance, influence may be seen as a trust relationship between influencers and their followers.

\subsection{Time aware recommender systems}

Most recommender systems that take temporal aspects into account are based on concept drift: older information is less important than recent information for predicting future user purchases. For this reason, \cite{Ding2005} proposed the use of the time-weight decay functions we used in this paper, in order to assign greater weight to the most recent ratings in similarity computations. In addition, \cite{gaillard2015time} propose a incremental matrix completion method, that automatically allows the factors related to both users and items to adapt "on-line" to concept drift hypothesis. Going further, \cite{liu2010online} propose an online incremental CF in which a decay function is used for similarity computations and another one is used for rating prediction. Time-weight functions are also used in other studies as in \cite{koren2009collaborative, karahodvza2015temporal, nzeko2017time}. 

Other approaches to concept drift assume that the importance of information used for recommendations is ephemeral, as in \cite{lathia2009temporal} where time is divided into slices and data is used only within a single slice. Such recommender systems therefore focus on user short-term preferences. It however seems that some preferences are stable and persist over time, and so that old information should also be included. For this reason, some works \cite{xiang2010temporal,li2007dynamic} capture both short-term preferences and long-term preferences and combine them in the recommendation process. For example, \cite{xiang2010temporal} propose STG to incorporate temporal aspects by separately modeling long-term preferences and short-term preferences within a graph model. 

\subsection{Content-based recommender systems}

These systems aim at recommending items similar to the ones the user liked in the past. A way to achieve this, developed in \cite{lops2011content}, is to match features associated to user preferences with those of items. Then, recommendation is performed in three steps: extracting relevant features from items, build user preference profiles based on item features, and finally select new items that fit user preferences. This approach is used in several domains such as recommendation of books \cite{mooney2000content} and recommendation of web pages \cite{pazzani1996syskill}.

Using content-based features may improve CF techniques by allowing more details on user favorite item features and increase the possibility to reach items that have not been selected in the past by other users. Some works \cite{balabanovic1997fab,basu1998recommendation,burke2002hybrid} indeed show that these hybrid recommender systems solve weaknesses of both approaches.

Recent work on content-based approaches are dedicated to the Social Book Search (SBS). The SBS Lab investigates book search in scenarios where users search with more than just a query, and look for more than objective metadata. It has two tracks. The first one is a Suggestion Track aiming at developing test collections for evaluating ranking effectiveness of book retrieval and recommender systems. The second one is an Interactive Track aimed at developing user interfaces that support users through each stage during complex search tasks and to investigate how users exploit professional metadata and user-generated content \cite{koolen2015overview}.

\subsection{Graph-based recommender systems}

The simplest graph-based recommender system rely on the classical bipartite graph (BIP) in which only user-item links are used. Most used algorithms are based on random walk \cite{baluja2008video}, like Injected Preference Fusion \cite{xiang2010temporal} and PageRank which is used in this paper; they compute a probability to reach items from the user under concern, and recommend the ones with highest probability.

Graph-based systems may be seen as CF systems, and so one may use the same idea as in hybrid recommender systems to improve them \cite{burke2002hybrid}. \cite{phuong2008graph} achieve this by adding a third node type: content nodes. The resulting graph ignores temporal aspects, though. To improve this, \cite{yu2014topic} propose the Topic-STG which incorporate content-based features and the temporal dynamic of STG. However these graphs handle each link regardless of its age, which contradicts the concept drift assumption. This is why we (\cite{nzeko2017time}) propose the Time-weight and content-based STG, where old links have a lower weight than recent ones. Up to our konwledge, none of these graph-based works takes advantage of content-based, time and trust information simultaneously.

We note that, despite the fact that recommender graphs are not much studied compared to model-based techniques such as matrix factorization or neural networks, they remain relevant. For example Pixie recommender system proposed by \cite{eksombatchai2018pixie} is the recent scalable graph-based real-time system developed and deployed at Pinterest. Given a set of user-specific pins as a query, Pixie selects in real-time from billions of possible pins that are most related to the query. To generate recommendations, Eksombatchai et al. develop Pixie Random Walk algorithm that uses the Pinterest object graph of 3 billion nodes and 17 billion edges. This has been made possible thanks to the technological evolution of Random Access Memories.

\section*{Conclusion}
\label{sec:conclusion}

Our main goal with this paper was to show that including several side information improves the quality of recommender graphs built for top-N recommendation task. For this purpose, we designed and implemented GraFC2T2, a recommender graph framework which makes it easy to explore various approaches for modeling and combining many features of interests for recommendation. In particular, GraFC2T2 extends classical bipartite graphs, session-based temporal graphs and link stream graphs by integrating content-based features, time-weight functions, and user trust into a personalized PageRank system.

The experiments we conducted on Epinions and Ciao datasets with F1-score, Hit ratio and MAP evaluation metrics show that best performances are always reached by graphs that integrate at least two side information and that graphs with time-weight always outperform the others. The resulting improvements are of at least 41\%. Moreover, comparison with state-of-the-art matrix factorization and classical user-based and item-based collaborative filtering methods confirms the relevance of GraFC2T2 framework for top-N recommendation. Good improvements obtained in recommender graphs by integration of side information do not guarantee such improvement for other types of recommender systems such as matrix factorization and neural network. We therefore consider inclusion of content-based, time and trust information simultaneously in such system as a key perspective.

\bibliographystyle{jimis-en}
\bibliography{nzekon-agr-jimis}

\appendix\footnotesize

\section{Acknowledgements}
We thank Rapha\"el Fournier, Tiphaine Viard and JIMIS reviewers for their helpful comments on previous versions. 
This work is funded in part by the African Center of Excellence in Information and Communication Technologies (CETIC), the Sorbonne University-IRD PDI program, and by the ANR (French National Agency of Research) under grant ANR-15-CE38-0001 (AlgoDiv).

\section{Appendix}
\label{sec:appendix}

In this section, we present the results obtained for top-20, -50 and -100. The section is divided in two parts: the first one presents performances obtained for all combinations of side information and basic graphs of the framework; the second highlights the 3 best combinations, according to basic graph and evaluation metric.

These two parts confirm observations made on top-10 results in the Section \ref{sec:experiments_results}. For example, recommender graphs that integrate simultaneously content-based, users' preferences temporal dynamic and trust relationship between users, are usually the best. Thus, we recommend the simultaneous integration of these three side information in order to increase the chances to achieve good performances.

\newpage

\begin{table}
\centering
\caption{Epinions Dataset - Performances with optimal settings for Top-20.}
\begin{tabular}{c}
\includegraphics[height=0.65\textwidth,width=0.7\textwidth]{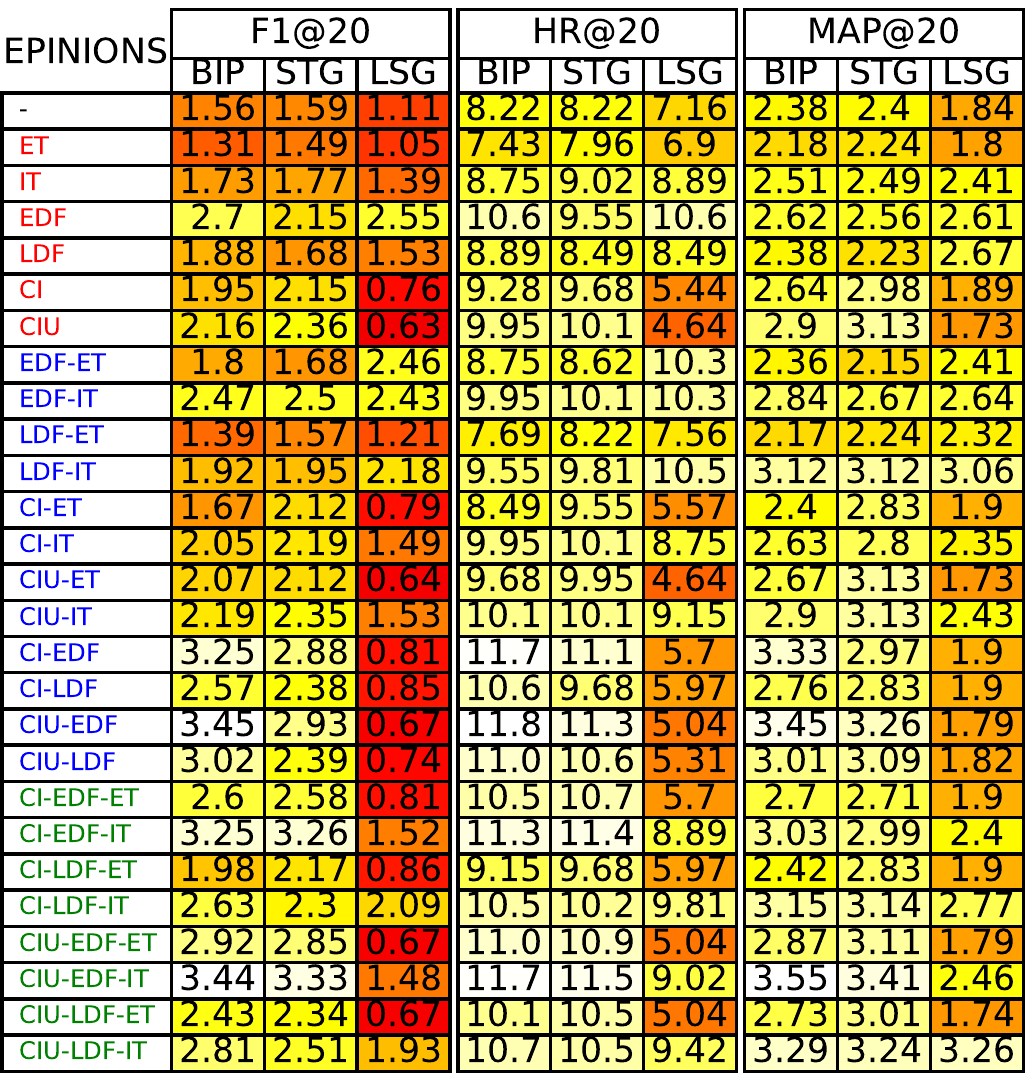}
\end{tabular}
\end{table}

\begin{table}
\centering
\caption{Epinions Dataset - Performances with optimal settings for Top-50.}
\begin{tabular}{c}
\includegraphics[height=0.65\textwidth,width=0.7\textwidth]{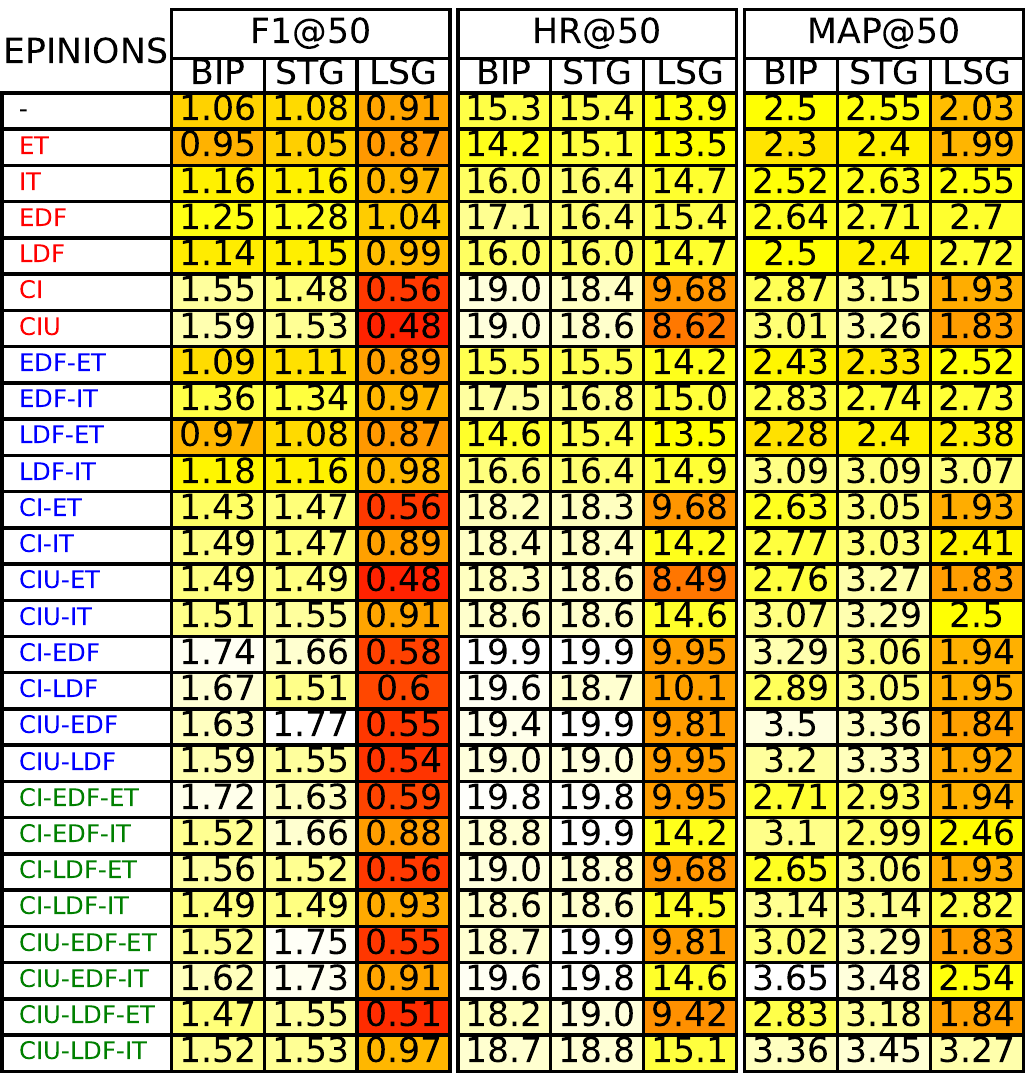}
\end{tabular}
\end{table}

\begin{table}
\centering
\caption{Epinions Dataset - Performances with optimal settings for Top-100.}
\begin{tabular}{c}
\includegraphics[height=0.65\textwidth,width=0.7\textwidth]{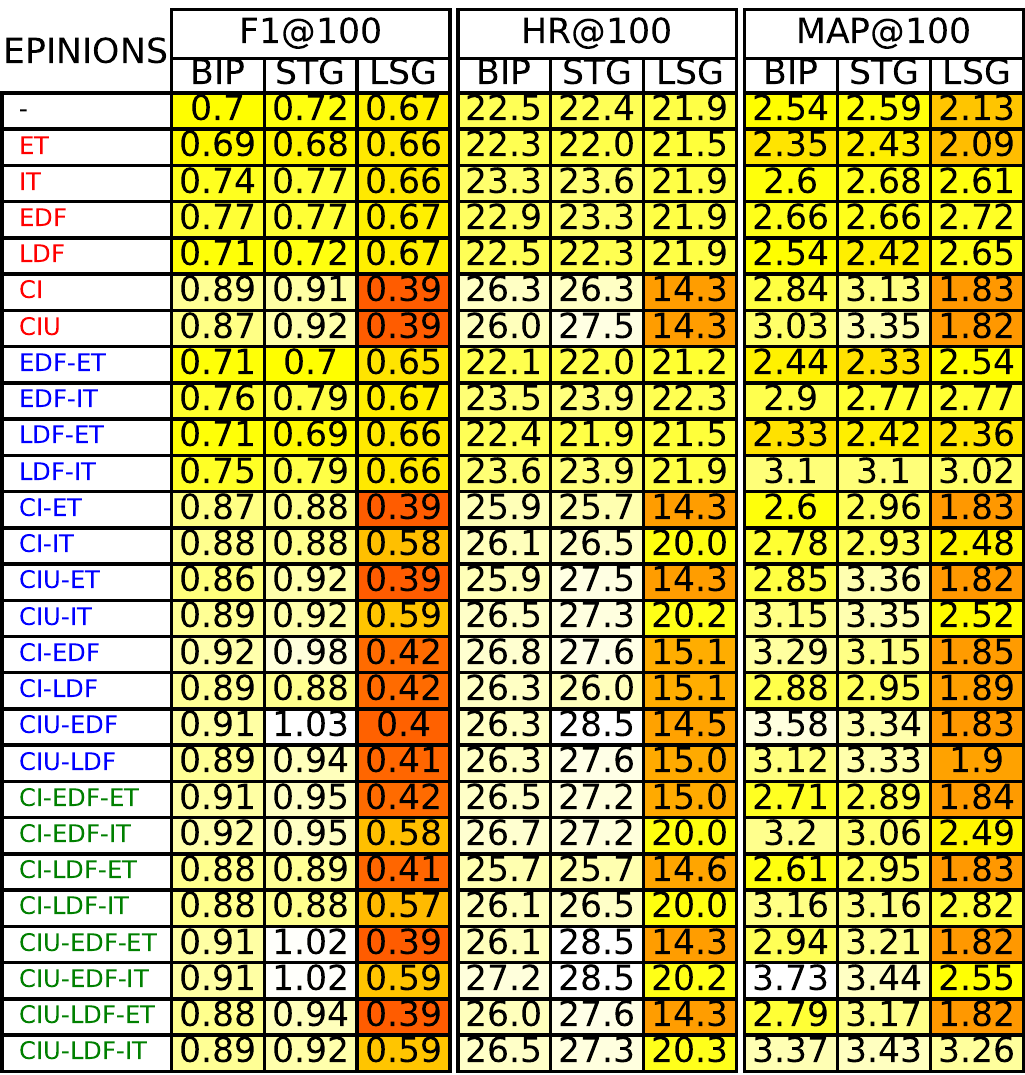}
\end{tabular}
\end{table}

\begin{table}
\centering
\caption{Ciao Dataset - Performances with optimal settings for Top-20.}
\begin{tabular}{c}
\includegraphics[height=0.65\textwidth,width=0.7\textwidth]{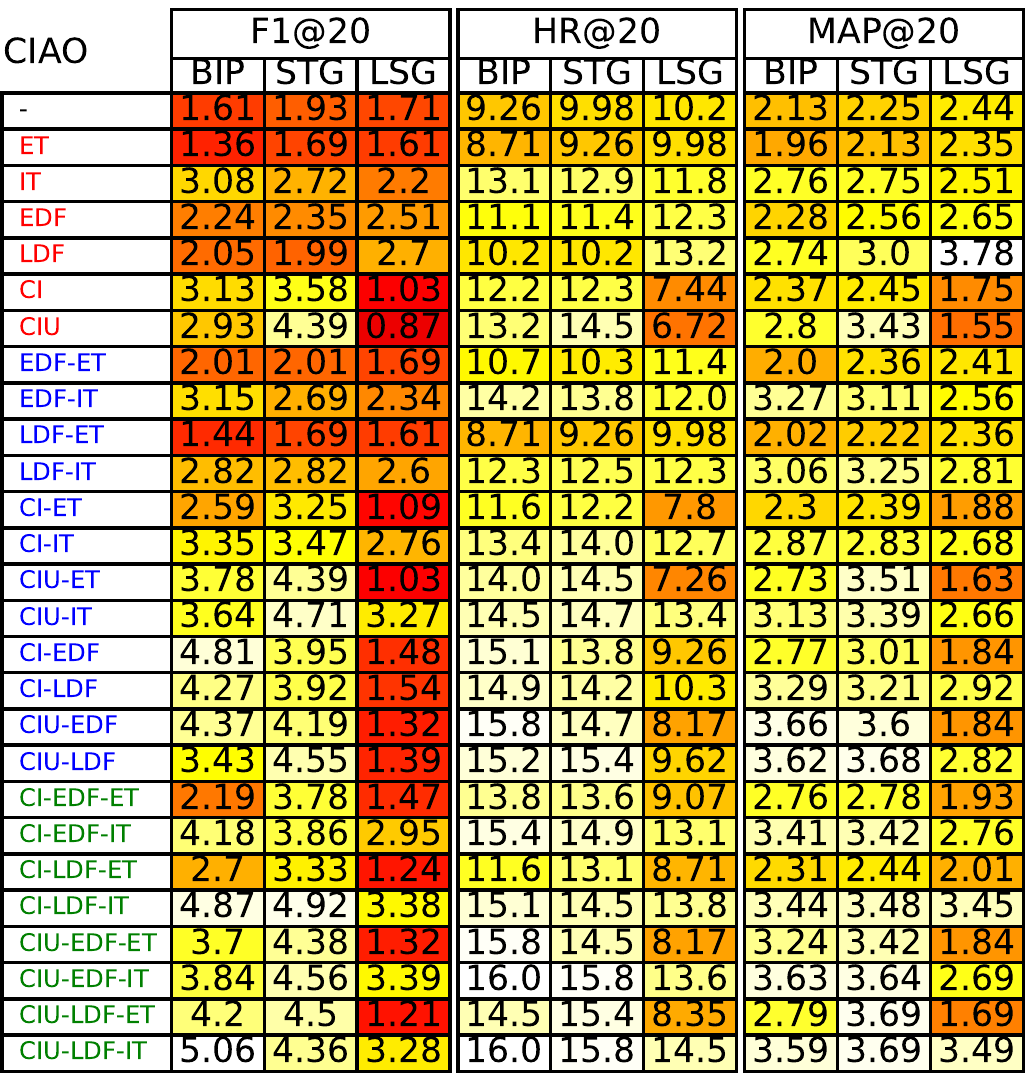}
\end{tabular}
\end{table}

\begin{table}
\centering
\caption{Ciao Dataset - Performances with optimal settings for Top-50.}
\begin{tabular}{c}
\includegraphics[height=0.65\textwidth,width=0.7\textwidth]{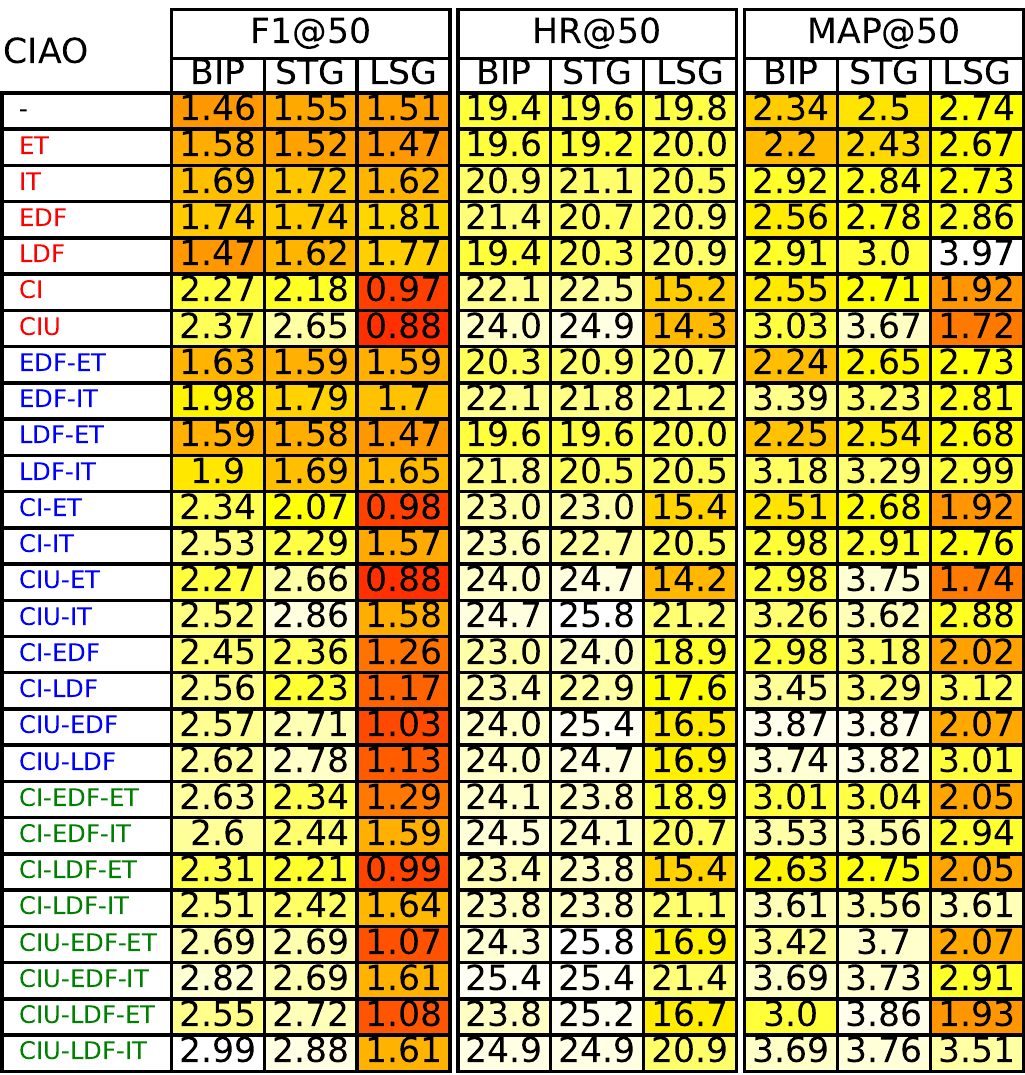}
\end{tabular}
\end{table}

\begin{table}
\centering
\caption{Ciao Dataset - Performances with optimal settings for Top-100.}
\begin{tabular}{c}
\includegraphics[height=0.65\textwidth,width=0.7\textwidth]{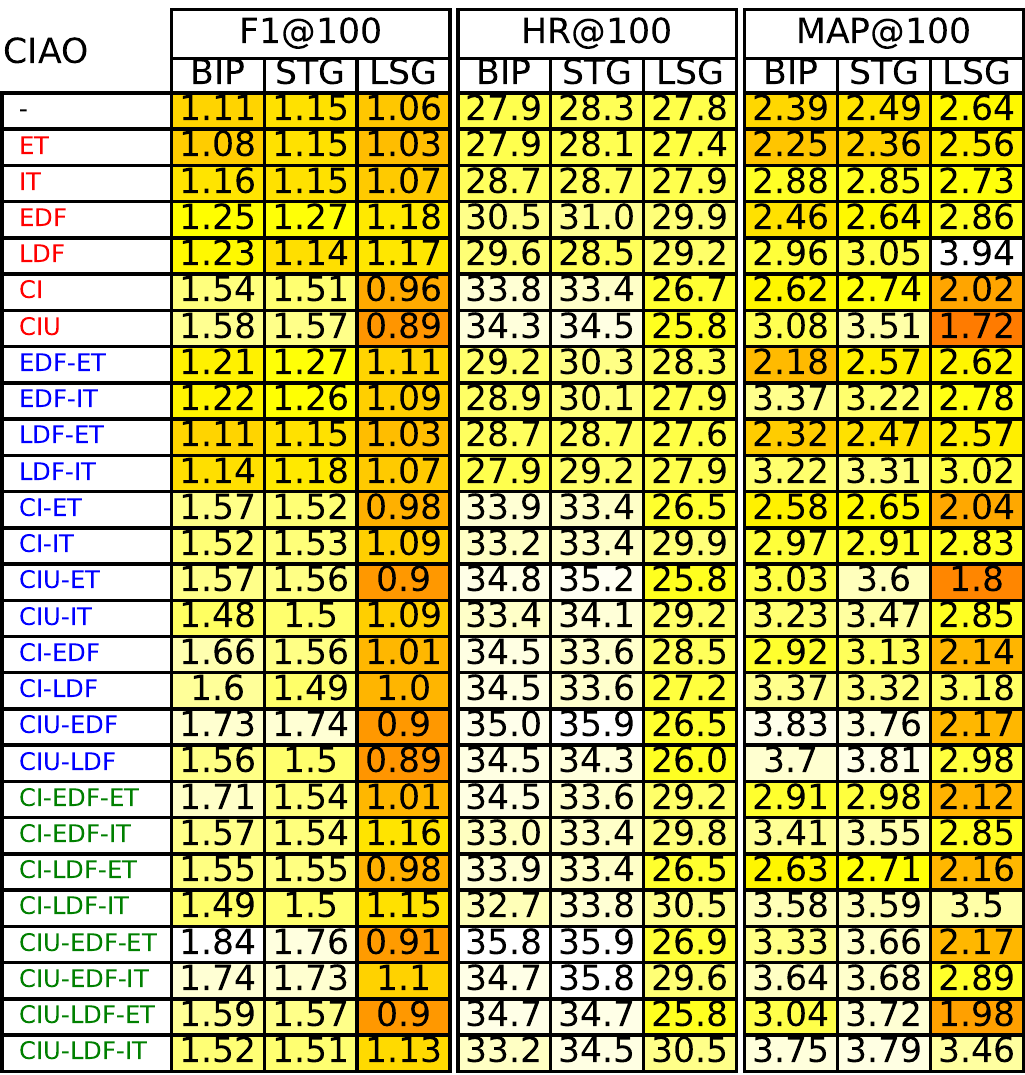}
\end{tabular}
\end{table}

\begin{table} [h]
\centering
\caption{Epinions Dataset - Best recommender graphs for Top-20, -50 and -100. Comparison of the three best recommender graph combinations with the associated basic graph.}
\label{tab_imp_all_epinions}
\begin{tabular}{c}
\includegraphics[height=0.32\textwidth,width=0.95\textwidth]{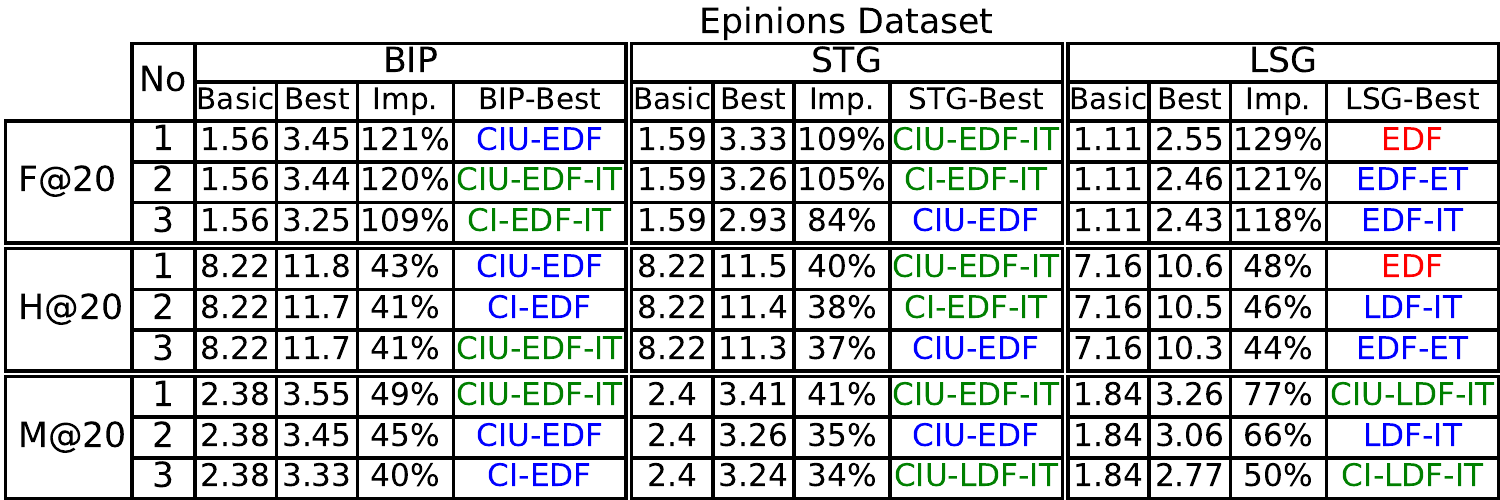}
\end{tabular}
\begin{tabular}{c}
\includegraphics[height=0.32\textwidth,width=0.95\textwidth]{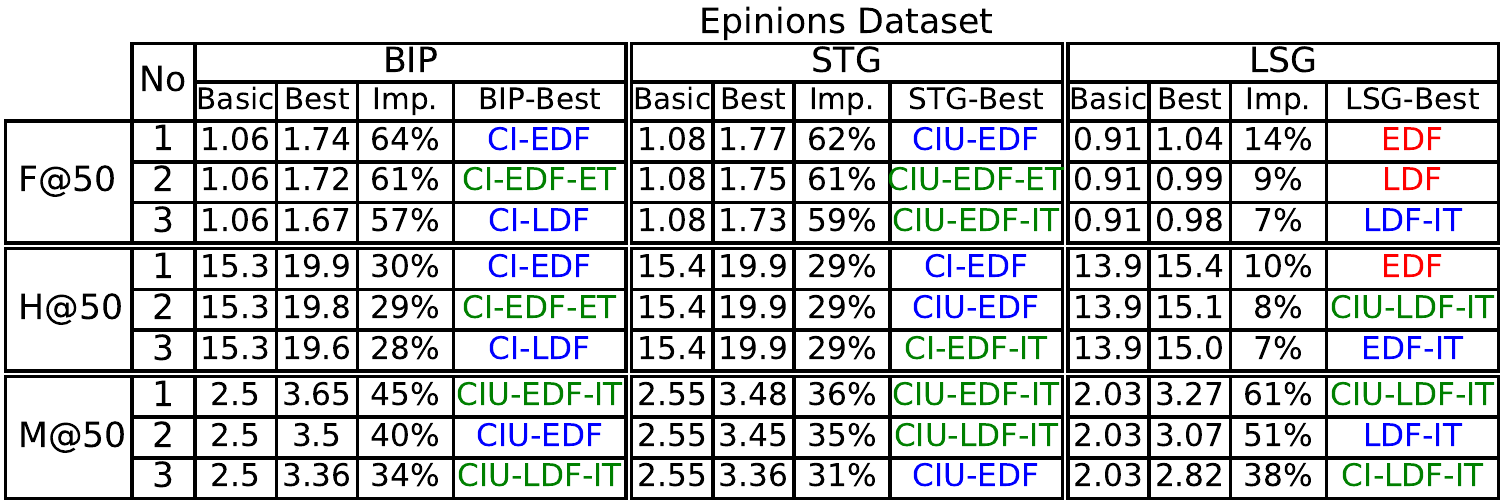}
\end{tabular}
\begin{tabular}{c}
\includegraphics[height=0.32\textwidth,width=0.95\textwidth]{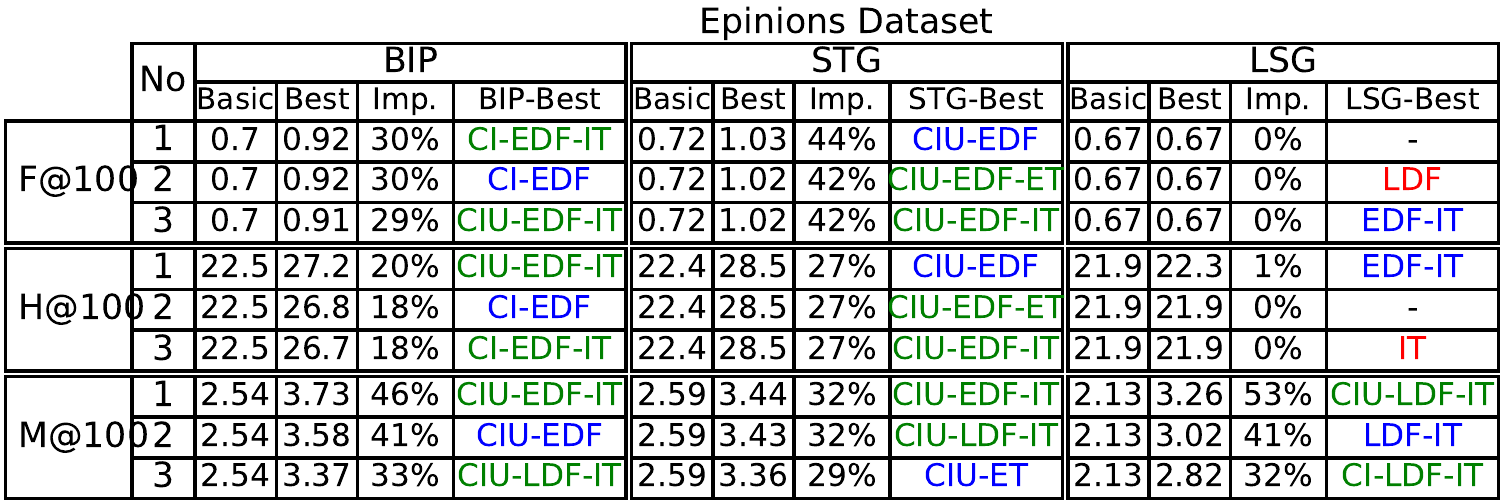}
\end{tabular}
\end{table}

\begin{table} 
\centering
\caption{Ciao Dataset - Best recommender graphs for Top-20, -50 and -100. Comparison of the three best recommender graph combinations with the associated basic graph.}
\label{tab_imp_all_ciao}
\begin{tabular}{c}
\includegraphics[height=0.32\textwidth,width=0.95\textwidth]{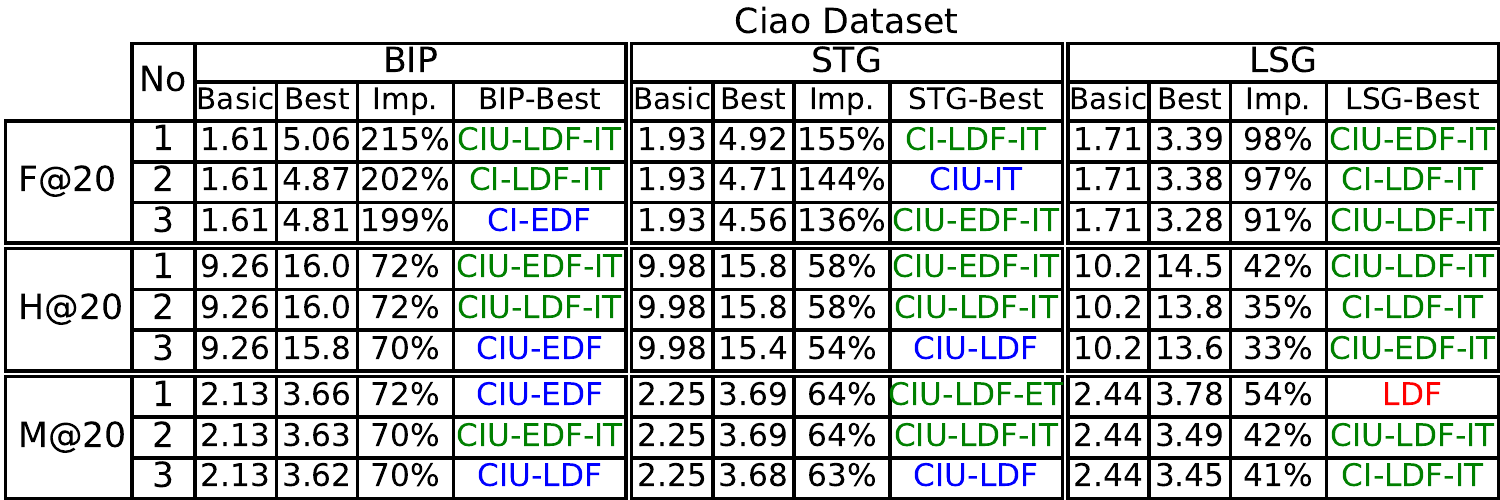}
\end{tabular}
\begin{tabular}{c}
\includegraphics[height=0.32\textwidth,width=0.95\textwidth]{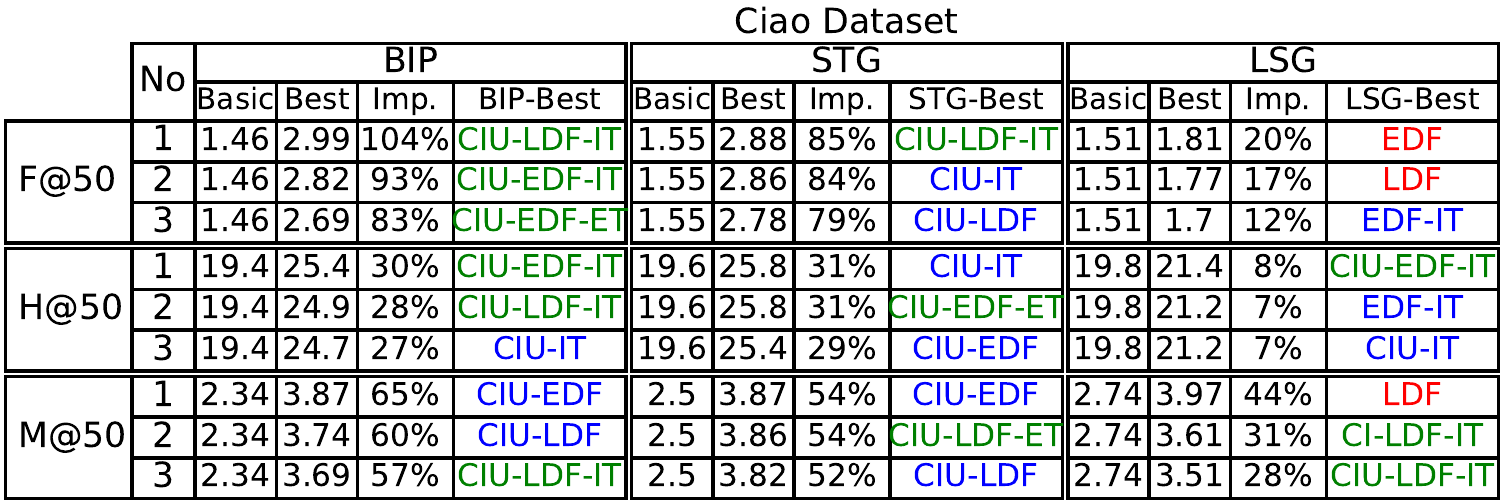}
\end{tabular}
\begin{tabular}{c}
\includegraphics[height=0.32\textwidth,width=0.95\textwidth]{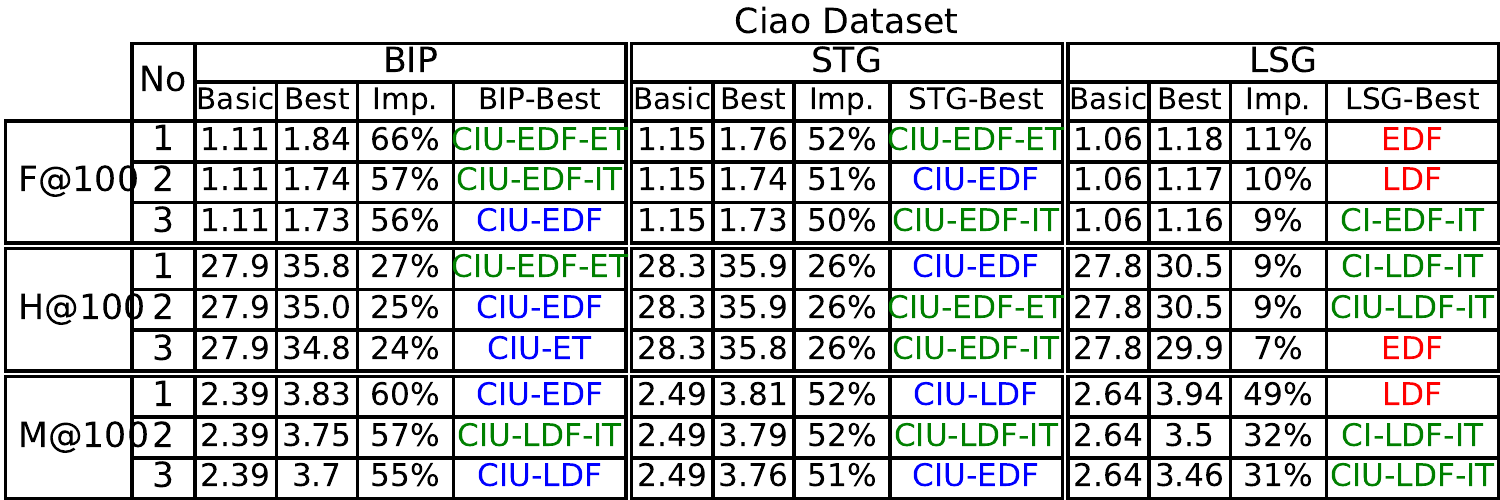}
\end{tabular}
\end{table}

\end{document}